\documentclass[conference]{IEEEtran}
\IEEEoverridecommandlockouts
\usepackage{cite}
\usepackage{amsmath,amssymb,amsfonts}
\usepackage{algorithmic}
\usepackage{graphicx}
\usepackage{textcomp}
\usepackage{xcolor}
\usepackage{etoolbox}

\usepackage{textcomp}
\usepackage{xcolor}
\usepackage{float}

\usepackage{xcolor,soul}
\usepackage{xspace}
\usepackage{enumitem}
\usepackage{titlecaps}
\usepackage{tabularx}
\usepackage{balance}
\usepackage{color}
\usepackage{comment}
\usepackage{graphicx}
\usepackage{grffile} 
\usepackage{subcaption}

\usepackage{url}
\usepackage{hyperref}

\def\BibTeX{{\rm B\kern-.05em{\sc i\kern-.025em b}\kern-.08em
    T\kern-.1667em\lower.7ex\hbox{E}\kern-.125emX}}

\newcommand{\sol}{{\em Camouflage}\xspace}
\newcommand{\solOne}{{\em Remodeling}\xspace}
\newcommand{\mix}{{\em Mixing}\xspace}
\newcommand{\hybrid}{{\em Hybrid}\xspace}
\newcommand{\eg}{{\em e.g.,}\ }
\newcommand{\ie}{{\em i.e.,}\ }

\begin{document}

\makeatletter
\def\ps@IEEEtitlepagestyle{%
  \def\@oddfoot{\mycopyrightnotice}%
  \def\@evenfoot{}%
}
\def\mycopyrightnotice{%
  {\footnotesize \textcolor{red}{\begin{tabular}[t]{@{}l@{}} This paper has been accepted for publication by the 2023 IEEE International Conference on Cloud Computing (CLOUD 2023). © 2023 IEEE. Personal use \\ of this material is permitted. Permission from IEEE must be obtained for all other uses, in any current or future media, including reprinting/republishing \\ this material for advertising or promotional purposes, creating new collective works, for resale or redistribution to servers or lists, or reuse of any copyrighted \\ component of this work in other works.\end{tabular}}}
  \gdef\mycopyrightnotice{}
}

\title{The Case for the Anonymization of Offloaded Computation}

\author{
\IEEEauthorblockN{Md Washik Al Azad}
\IEEEauthorblockA{University of Notre Dame\\
malazad@nd.edu}
\and
\IEEEauthorblockN{Shifat Sarwar}
\IEEEauthorblockA{University of Nebraska at Omaha \\
ssarwar@unomaha.edu}
\and
\IEEEauthorblockN{Sifat Ut Taki}
\IEEEauthorblockA{University of Notre Dame \\
staki@nd.edu}
\and
\IEEEauthorblockN{Spyridon Mastorakis}
\IEEEauthorblockA{University of Notre Dame \\
mastorakis@nd.edu}
}
\makeatletter
\patchcmd{\@maketitle}
  {\addvspace{0.5\baselineskip}\egroup}
  {\addvspace{-1.0\baselineskip}\egroup}
  {}
  {}
\makeatother

\setlength{\skip\footins}{4pt}

\maketitle

\begin{abstract}
Computation offloading (often to external computing resources over a network) has become a necessity for modern applications. At the same time, the proliferation of machine learning techniques has empowered malicious actors to use such techniques in order to breach the privacy of the execution process for offloaded computations. This can enable malicious actors to identify offloaded computations and infer their nature based on computation characteristics that they may have access to even if they do not have direct access to the computation code. In this paper, we first demonstrate that even non-sophisticated machine learning algorithms can accurately identify offloaded computations. We then explore the design space of anonymizing offloaded computations through the realization of a framework, called \sol. \sol features practical mechanisms to conceal characteristics related to the execution of computations, which can be used by malicious actors to identify computations and orchestrate further attacks based on identified computations. Our evaluation demonstrated that \sol can impede the ability of malicious actors to identify executed computations by up to 60\%, while incurring modest overheads for the anonymization of computations.
\end{abstract}

\begin{IEEEkeywords}
Computation Anonymization, Machine Learning, Computation Offloading, Cloud Computing, Computation Graphs
\end{IEEEkeywords}

\section {Introduction}
\label{sec:intro}


Offloading computation, often to external computing resources over a network, has become a necessity for modern applications, which typically require the execution of compute-intensive tasks~\cite{mach2017mobile}. Nevertheless, offloading computation in untrusted and/or compromised environments, such as compromised Virtual Machine (VM) instances in cloud environments, remains a major challenge~\cite{boneh2015hosting}. 

At the same time, advanced technologies, such as machine learning,
amplify the capabilities of attackers and offer new opportunities to breach the privacy of the execution process of offloaded computations. This can enable attackers to identify offloaded computations and infer their nature based on computation characteristics that attackers may have access to (\eg completion time, CPU usage, size and/or number of inputs/outputs) even if they do not have direct access to the computation code~\cite{mazloom2018secure, ohrimenko2015observing}. Identifying sensitive executed computations 
can lead to the orchestration of more sophisticated attacks by malicious actors (e.g., to gain access to encryption/decryption keys~\cite{10.1145/2382196.2382230, aciiccmez2007cache}).  


To address this challenge, in this paper, we explore the design space and different practical mechanisms to anonymize offloaded computations in 
environments, where attackers may attempt to identify executed computations based on computation characteristics they have access to. We realize these mechanisms in the context of a framework, which we call \sol.
\sol aims to conceal characteristics of the execution of computations in different ways: (i) by introducing ``randomness'' to the computation process, while adding modest overheads and without impacting the computation execution correctness; (ii) by making independent computations ``look alike''; and (iii) by combining the execution of multiple independent computations that are concurrently running on a system to conceal properties of the execution of individual computations. The contributions of our work are the following:

\begin{itemize}

\item We present practical mechanisms to anonymize computations aiming to hamper the ability of attackers that use machine learning techniques to identify executed computations. These mechanisms are realized as different modes of operation of the \sol design. 

\item We implement a \sol prototype, which we evaluate based on both synthetic and real-world datasets of computations. Our evaluation results demonstrate that \sol is able to decrease the ability of attackers, who take advantage of machine learning algorithms for the identification of executed computations, by up to 60\%. \sol introduces modest overheads to the execution process of computations for their anonymization.

\end{itemize}




\section{Use Cases and Motivation}
\label{sec:use-cases}

\subsection {Use Cases}

Let us consider a cloud or edge computing scenario where users offload computational tasks for execution to VM instances, which may be compromised by an attacker. For instance, a VM instance may get compromised when a user installs or updates a program that is considered to be legitimate and harmless but contains a Remote Access Trojan (RAT)~\cite{rezaeirad2018schrodinger, 9229824}. In this context, an attacker can perform limited activities on the compromised system, such as sending or receiving data between the attacker and the victim system, retrieving the list of running programs on the system, and monitoring the utilized computing resources by each program. Let us also assume that an attacker does not have privileges to run or modify programs on the compromised system. To infer what 
computations are performed on the system, the attacker can create a behavior profile of the computations running on the system with limited access and information. This allows an attacker to identify when a user performs certain (potentially sensitive) computations, such as user authentication, encryption/decryption of data, processing of confidential data (\eg patients' medical data), and orchestrate attacks in a more targeted and timely manner based on the inferred computation types~\cite{10.1145/2382196.2382230, aciiccmez2007cache}. 

In this context, let us consider a scenario where a ride-sharing service running on compromised VMs in a cloud environment has multiple sub-processes. For example, the service requires separate processes for communicating with riders and drivers, payment processing, GPS tracking, maps and traffic congestion updates, finding the fastest route, and estimating the cost of a trip. If an attacker has limited access to such VM instances, the attacker can gather information about performed computations on each instance. This enables the attacker to identify in which instances sensitive tasks (\eg payment processing) are performed and deploy further attacks, such as Distributed Denial of Service (DDoS) attacks~\cite{agrawal2019defense}, to these instances to negatively impact the whole service. Furthermore, identifying the available processes can lead to revealing critical business insights, such as the number of active users by monitoring the communication process with client applications and the number of trips per day by analyzing the number of instances of the fastest route finder process.


Let us consider another scenario where a user would like to run a Deep Neural Network (DNN) model for a computer vision service on a third-party 
cloud VM instance. The DNN model is the user's intellectual property and the model architecture, such as the number of layers and the size of each layer, are critical information. This information needs to be protected in order to preserve the intellectual property of the user and to prevent attacks (\eg adversarial machine learning) against the DNN model, which could degrade its performance. Prior research has demonstrated that it is possible to extract the complete model architecture based on resource utilization information, in some cases even without any prior knowledge about the targeted model~\cite{hu2020deepsniffer}. An 
adversary can deploy such a model extraction attack to gain access to a DNN model architecture and other parameters related to this model.

\subsection {Motivation}

Based on the use cases presented above, attackers can gather information about processes and infer what computations users perform. Attackers can also create a profile of the computations running on targeted systems, leading to privacy violations and security issues. 
Such attacks can leak sensitive 
information (\eg as discussed above for the ride-sharing application use case), which may lead to privacy violations and financial loss, or offer advantages to business competitors. The proliferation of machine learning techniques amplifies the capabilities of attackers and offers a tool that attackers can take advantage of to realize use cases like the ones we discussed above, even if they are not machine learning experts.

\begin{figure}[!t]
 \centering
 \includegraphics[scale=0.33]{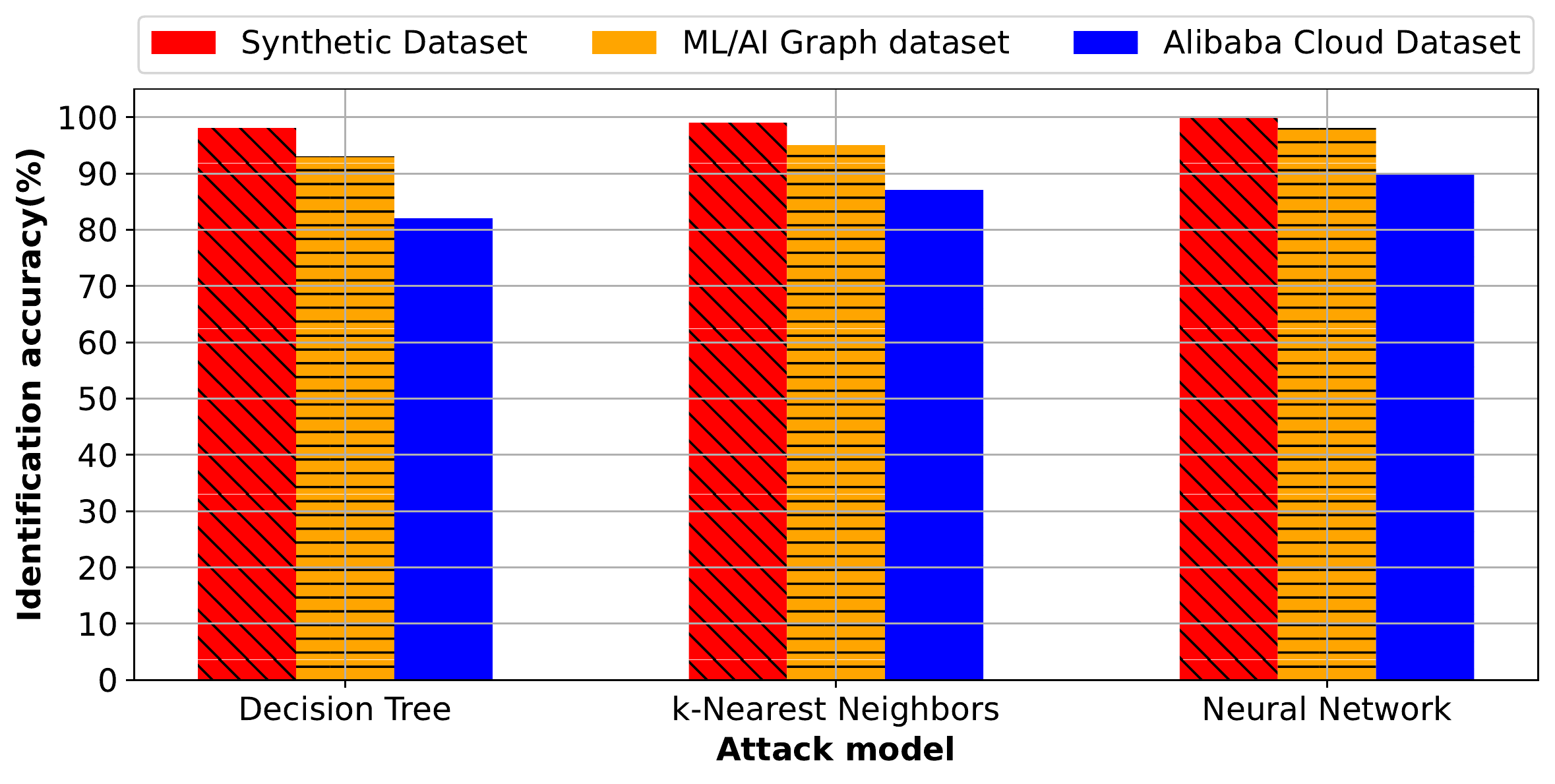}
 \vspace{-0.2cm}
 \caption{Computation identification accuracy for different ``out-of-the-box'' machine learning models and datasets.} 
 \label{Figure:attack_acc_before}
 \vspace{-0.5cm}
\end{figure}

To verify our intuition, we ran experiments with different ``out-of-the-box'' machine learning models and trained them to identify computations in a black-box fashion (\ie without having direct access to the internal structure of computations, but rather based on characteristics collected during the execution of computations, such as the number and size of inputs and outputs, CPU usage, and time to completion). We used three datasets of computations: (i) a synthetic dataset of computations; (ii) a Machine Learning (ML)/Artificial Intelligence (AI) graph dataset; and (iii) and a real-world dataset of computation services offered by Alibaba cloud. We provide further details about these datasets in Section \ref{sec:eval}. As Figure \ref{Figure:attack_acc_before} shows, in general, we are able to identify computation processes based on their characteristics (without access to their code) with more than 80\% accuracy, and, in some cases, the accuracy reaches 99\%. To address these issues, in this paper, we present \sol, a computation anonymization framework, which allows users to offload computations for execution, while impairing the attackers' ability to identify the executed computations.


\section{Threat Model}
\label{sec:threat}



\noindent\textbf{Access to a targeted system:} We consider a scenario where the targeted system is infected by a malware, such as a Remote Access Trojan (RAT)~\cite{rezaeirad2018schrodinger} or a malicious rootkit, that allows the attacker to monitor the victim system activities in a limited fashion. We assume that targeted systems will faithfully execute computations and will not attempt to produce bogus computation results. We further assume that users run computations as programs and attackers do not have access to the source code of these programs, but they can have access to characteristics (features) of such programs. Nevertheless, we assume that it will be rather expensive for attackers (in terms of time and effort) to reverse engineer these programs.

\noindent\textbf{Feature collection:} Once an attacker has access to a targeted system, they start collecting information about the running processes on the system. The collected information may include CPU and memory usage per process, the run (completion) time of each process, the size of input and output data, and network activity, among others. Subsequently, the attacker can use this information to identify the computation processes running on the targeted system using machine learning models. 


\noindent\textbf{Identifying computations with an ``attack model'':} After collecting the features of a computation, the attacker will use a trained machine learning model (we call it an ``attack model'') to identify the computation processes executed by the targeted user. To train the model, we consider that the attacker can create a dataset of different types of computation processes and use this dataset to train this model, which can be used to identify the computation processes executed on behalf of targeted users. To create such a dataset, 
an attacker can 
execute (in isolation from users) computation services (with various configuration and input parameters) that may be offered to users and collect features about the execution of these services.

\noindent\textbf{Adaptive attack models:} An attacker may realize that anonymization techniques are deployed to hide the identity of computations. As a result, the performance of an attack model can degrade over time and an attacker may attempt to adapt to these techniques. To achieve that, we assume that an attacker can repeatedly re-train an attack model over time to maintain its effectiveness. 



\section{\sol Design}
\label{sec:design}


\subsection{Design Assumptions and Overview}

\begin{figure}[!t]
 \centering
 \includegraphics[width=1\columnwidth]{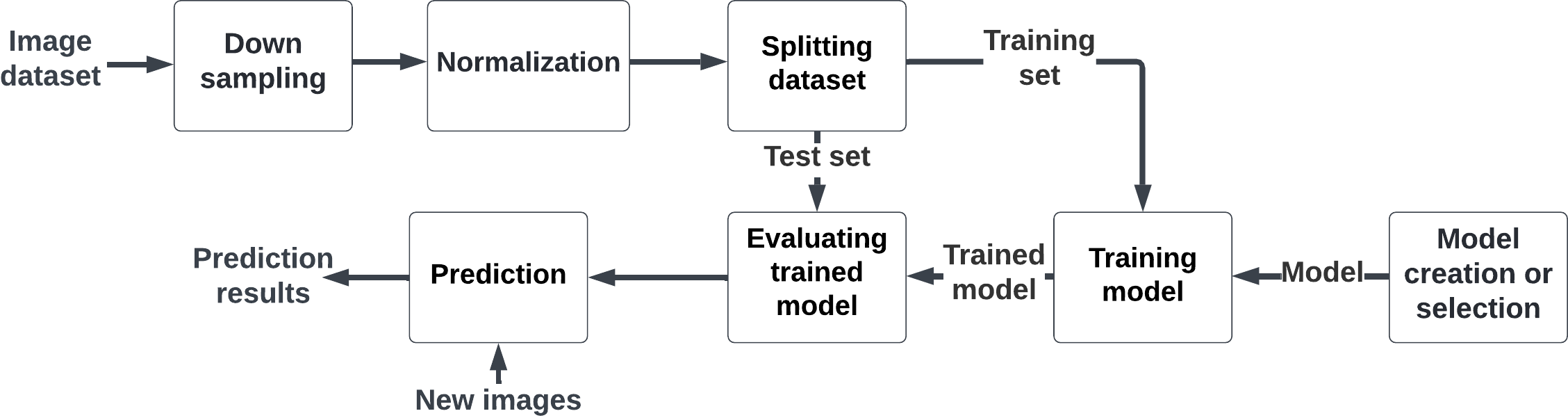}
 \caption{Example of a computation graph representing the process of training a deep learning model.} 
 \label{Figure:DAG}
 \vspace{-0.5cm}
\end{figure}

We assume that computations are formulated as Directed Acyclic Graphs (DAGs) \cite{bang2008digraphs}, which indicate the overall computation workflow. Each node of a DAG represents a specific part or step of the overall computation process (\eg a function) and the input(s) and output(s) of each node represent the edges of a DAG. Each node accepts one or more inputs and yields one or more outputs. We also consider that the inputs and the outputs can be either encrypted or unencrypted. Let us consider the process of training a deep learning model with an image dataset as a computation workflow (Figure~\ref{Figure:DAG}). This process involves multiple steps (functions), such as down-sampling and normalizing images, splitting the image dataset into training and testing sets, creating/selecting a deep learning model, feeding the training sets to the model and training the model, as well as measuring the performance of the trained model with test sets. Each of these steps can be represented as a node which accepts the output(s) of the previous step(s) as its input(s). In other words, the execution of a part of the overall computation represented through a specific node depends on the completion of the computation parts represented through previous nodes of the DAG. 

\sol aims to anonymize computations based on three modes of operation: \solOne, \mix, and \hybrid. \solOne conceals the computation process identity by 
changing the characteristics of computations to confuse attackers. 
\mix takes a different approaches where the nodes of multiple, simultaneously-running, independent computation graphs on a system are mixed to hide the identity of the executed computations from attackers. \mix is feasible when there are multiple concurrent computations to be executed, while \solOne can be used to anonymize standalone computations regardless of the availability (or not) of concurrently running computations. Finally, \hybrid combines the \solOne and \mix modes to first conceal the identity of computations and then mix multiple anonymized computations. 

\subsection{\solOne Mode}
\label{subsec:solone}

\begin{figure}[t!]
 \centering
 \includegraphics[width=0.9\columnwidth]{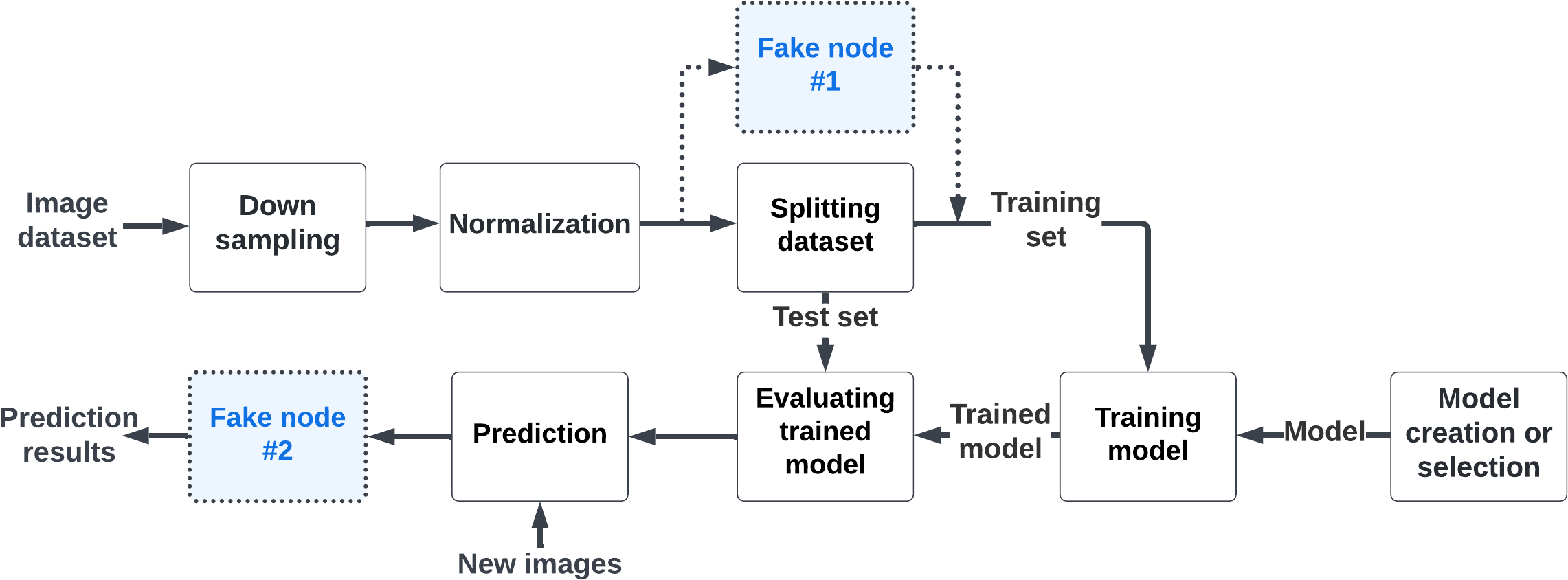}
 \caption{Example of a computation graph representing the process of training a deep learning model after the addition of ``fake'' nodes.} 
 \label{Figure:DAG-fake}
 \vspace{-0.5cm}
\end{figure}

In this mode, \sol aims to conceal computation features that an attacker can potentially exploit to identify computation processes. Let us consider hashing algorithms as an example. Such algorithms have in general fixed (and well known) output sizes, which are independent of the sizes of input data. In a similar manner, the input and output sizes of sorting algorithms are the same. In other words, specific types of computation processes are likely to be correlated with certain features, which can be exploited by attackers to infer certain computation processes. To conceal computation features, the \solOne mode of \sol employs mechanisms that: (i) introduce ``randomness'' to the computation process, while adding modest overheads to this process and without impacting its correctness; and (ii) make different computation processes ``look alike''. These mechanisms broadly fall under one of the following categories:

\begin {itemize}[leftmargin=0cm,itemindent=.3cm,labelwidth=\itemindent,labelsep=0cm,align=left, noitemsep, topsep=0pt]


\item \emph{Deliberate manipulation of inputs and outputs:} This category includes mechanisms that alter either the number or the size of inputs and outputs. This can be achieved by introducing additional (unused) inputs and/or outputs or by padding inputs and/or outputs respectively. 

\item \emph{Deliberate manipulation of the computation process:} This category includes mechanisms that alter the structure and characteristics of computation processes. This can be achieved by introducing additional parts to the overall computation process, which alter the execution times of computations and the utilization of computing resources (\eg CPU and memory usage). For example, these additional parts can be represented as ``fake'' nodes added to computation graphs (Figure~\ref{Figure:DAG-fake}). These nodes are innocuous in the sense that they do not impact the correctness of the computation process, but they operate as every other node in a computation graph by requiring a certain amount of time and resources (\eg CPU and memory) to be completed. As indicated in our evaluation results in Section~\ref{subsec:results}, the more ``fake'' nodes added to computation graphs, the more challenging it becomes for attackers to identify computation processes, however, the overhead of the computation process may increase.

\end{itemize}

\subsubsection{\solOne Mode Operation Workflow} 

\begin{figure}[t!]
 \centering
 \includegraphics[width=1\columnwidth]{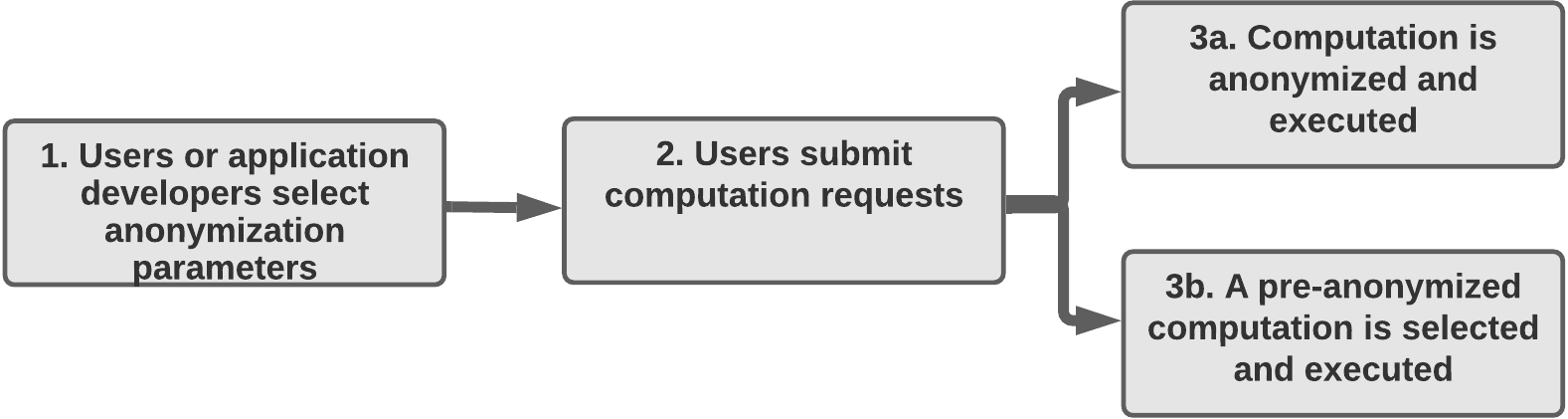}
 \caption{The operation workflow of the \solOne mode.} 
 \label{Figure:workflow}
 \vspace{-0.5cm}
\end{figure}

In Figure~\ref{Figure:workflow}, we present the operation workflow of the \solOne mode. It consists of three steps, which we describe below.

\noindent \textbf{Step 1:} \sol offers flexibility to users and application developers to determine the anonymization parameters. Such parameters may include which computation features to conceal and to what extent users would like to anonymize their computation processes depending on the sensitivity of these processes as well as constraints in terms of latency and resources. We call this extent ``anonymization level''. As the anonymization level increases, \sol will, for example, add more fake nodes and/or increase the padding size for inputs and outputs. This step can either be an initialization step (performed once for all computations of the same kind) or performed once per computation request. 


\noindent \textbf{Step 2:} Once the features to be concealed and the anonymization level have been selected, users will submit one or more requests for computation. Such computation requests may be offloaded to remote computing resources or sent to local computing resources for execution.

\noindent \textbf{Step 3:} This step can take place in two different manners: (i) anonymize a received computation on the fly and execute it (step 3a); or (ii) execute the requested computation, which has been (pre-anonymized) anonymized in advance (step 3b). Step 3a serves cases where computations may not be known in advance (\eg users offload a certain computation graph to be executed when they submit a request for computation) or users select different anonymization parameters for each computation request. On the other hand, step 3b serves cases where the formulations of computation (computation graphs) may be available in advance (\eg a cloud server offering a set of pre-determined services) and the anonymization parameters remain the same. Let us consider a cloud service formulated as a computation graph $g$. In the context of step 3b, anonymized graphs $ag_1, ..., ag_n$ of $g$ are created in advance to avoid the overhead of anonymizing computations on the fly. Once a user computation request is received, an anonymized graph among $ag_1, ..., ag_n$ is selected and executed.

\subsection{\mix Mode}

In this mode, we aim to mix (combine) the execution of multiple computation processes in order to conceal features and properties of the execution of individual computations. The intuition behind the design of this mode is to take advantage of the heterogeneity of different computations that may run concurrently on a system as the means to anonymize the computation execution.  

\begin{figure}[t!]
 \centering
 \includegraphics[width=1\columnwidth]{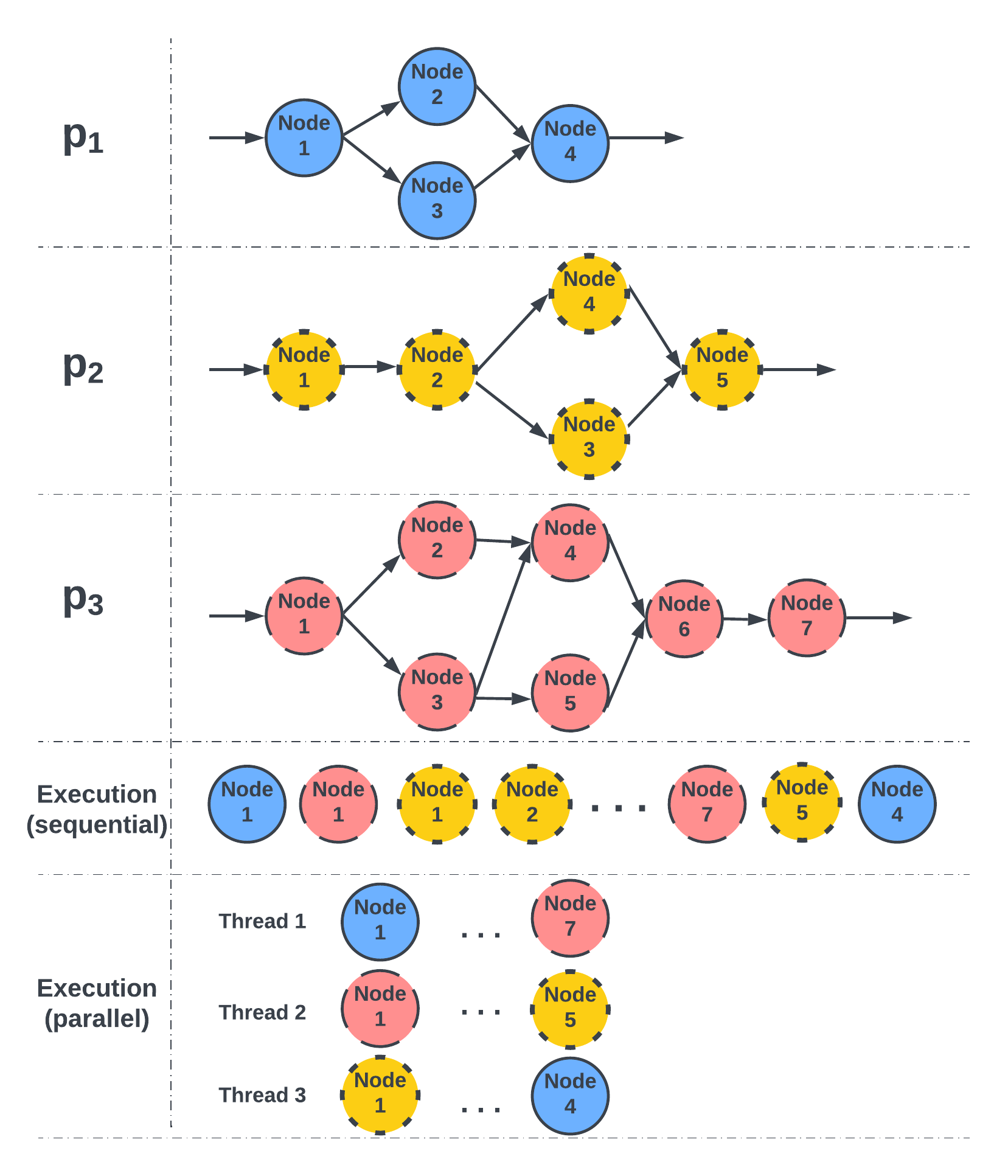}
 \vspace{-0.5cm}
 \caption{Example of mixing independent computation processes.} 
 \label{Figure:mix}
 \vspace{-0.5cm}
\end{figure}

We present an example of this mode in Figure~\ref{Figure:mix}, where three computation processes $p_1, p_2, p_3$ are mixed. In this example, the first node of the computation graph of $p_1$ is executed first, followed by the first node of the computation graph of $p_3$, the first node of the computation graph of $p_2$, the second node of the computation graph of $p_2$, and so on so forth (sequential \mix mode). During the operation of the \mix mode, we take into account the following considerations: 

\begin {itemize}[leftmargin=0cm,itemindent=.3cm,labelwidth=\itemindent,labelsep=0cm,align=left, noitemsep, topsep=0pt]

\item \emph{Heterogeneous computations:} Computations may be heterogeneous in nature in the sense that they may result in different usage of computing resources and may also require different amounts of time to complete. 

\item \emph{User/application constraints:} Applications may have different constraints in terms of latency/delay. Some applications may be delay-sensitive, thus requiring computations to be executed sooner than computations associated with delay-tolerant applications.

\end {itemize}

In other words, delay-sensitive computations could end up being completed after delay-tolerant computations. To address this challenge, \sol allows application developers or users to mark certain computations as delay-sensitive, so that their execution can be prioritized during the \mix mode. This prioritization is achieved through a weighted process for the selection of the next graph node to be executed. Specifically, the nodes of each graph to be mixed are assigned with a certain weight; the more delay-sensitive a computation process is, the greater the assigned weight will be. The next graph node to be executed is selected in a weighted random fashion. As a result, delay-sensitive computations will be prioritized (favored) to be completed first, while being mixed with other computations as the means to achieve anonymization.

Another parameter we consider in the context of the \mix mode is the number of computation processes that are being mixed and executed at the same time. We call this parameter ``batch size''. As our evaluation results in Section~\ref{subsec:results} demonstrate, the greater the value of the batch size is, the harder it becomes for attackers to identify computations. Greater batch size values, however, result in higher overheads during the mixing and execution of computations. 

In addition to the sequential \mix mode, mixed computations can be executed in parallel as we illustrate at the bottom of Figure~\ref{Figure:mix} (parallel \mix mode). In this case, multiple threads may be employed to execute nodes (parts) of the mixed computation graphs simultaneously. Parallelism reduces the total execution time of computations without impairing the ability of \sol to anonymize computations.


\subsection{\hybrid Mode}

In this mode, \sol first conceals computation features that an attacker could potentially exploit to identify computation processes and subsequently mixes these processes. This essentially acts as a hybrid mode of operation that combines both the \solOne and the \mix modes. In the same manner as in the \mix mode, the computations processes that are mixed can be executed in parallel to reduce the overall execution time. 

As demonstrated through our evaluation results (Section~\ref{sec:eval}), the \hybrid mode can further hamper the ability of attackers to identify computations. This comes at the cost of executing parts (nodes) of computation processes that could be ``fake'' (introduced for anonymization purposes) as compared to the the \mix mode, where all the mixed computation parts need to be actually executed.

\subsection{End-to-End \sol Framework}

The design of \sol as an end-to-end framework for the anonymization of offloaded computation includes the following components: (i) a \sol daemon running on user devices; and (ii) a \sol engine running on the computing resources (\eg servers) that execute computations on behalf of users. The user device daemon is an optional component and is used in cases where users would like to select a specific anonymization mode (\solOne, \mix, or \hybrid) and the parameters relevant to each mode. The \sol daemon is also responsible for establishing secure (SSL/TLS) connections with computing resources and offloading computations as needed by applications. 

The \sol engine is responsible for receiving and anonymizing computations offloaded by users. Computations may be offered as a service to users and may be pre-installed by service providers or application developers on computing resources. In such cases, users need to select the computation to be executed, provide the input parameters, and anonymization related parameters (if any). Users can also dynamically define new computations (different than the pre-installed ones) and offload them to computing resources for execution. Once computations are anonymized and executed, the output(s) of their execution are returned to user devices (if needed).
\section {Evaluation}
\label{sec:eval}



\subsection {\sol Application Programming Interface (API)}

We have implemented a prototype of \sol\footnote{We make our \sol implementation code and used datasets publicly available to the research community at \url{https://github.com/ShifatSarwar/ComputationAnonymization}.} which supports three modes (\solOne, \mix, and \hybrid). The API of \sol comes with four main functions, so that a user can create a computation graph and anonymize it. To create a computation graph, a user first calls the \verb|CreateGraph()| function, which creates a graph instance with no nodes and no edges. To add nodes and edges to the graph, \verb|ConnectNodes(functionName, *args)| can be used. \verb|ConnectNodes()| accepts a node name \verb|functionName=foo| and connects \verb|foo| with other graph nodes (\verb|*args|), on which \verb|foo|'s input(s) depend. Once all nodes are connected to form a computation graph, the graph needs to be compiled by using \verb|Compile()| to check that there are no violations of DAG conditions. Finally, the compiled graph can be anonymized by calling \verb|Anonymize(mode,*args)|. As an example, let us consider the following code snippet to anonymize the computation graph of training a deep learning model as shown in Figure \ref{Figure:DAG}.

{\small

\begin{verbatim}
def DownSampling(data_array):
    ...........
    return data_array

def Normalization(data_array):
    ...........
    return data_array
...........
...........
    
G = Camouflage.CreateGraph()
G.ConnectNodes(functionName="Normalization",\
                    "DownSampling")
...........
...........
G.Compile()
G.Anonymize(mode="Remodeling",level=2)
\end{verbatim}

}

In this example, a user implements each step of model training (\eg down sampling, normalization, dataset splitting) as an individual function. Then the user connects these functions (\eg the \verb|DownSampling()| function to the \verb|Normalization()| function) using the \sol API (\ie \verb|ConnectNodes()|). After connecting all functions and compiling the graph, the graph is anonymized through the \solOne mode with an anonymization level equal to 2.

\subsection {Evaluation Setup}

\begin{figure}[!t]
 \centering
 \includegraphics[width=1\columnwidth]{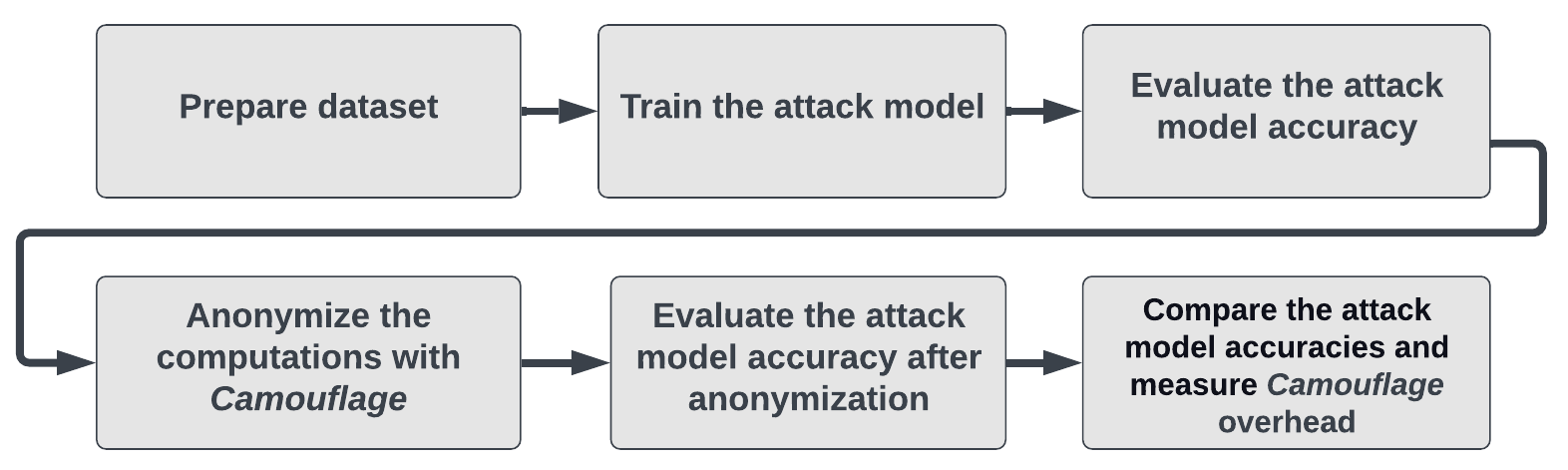}
 \vspace{-0.5cm}
 \caption{An overview of the \sol evaluation process.} 
 \label{Figure:eval_steps}
 \vspace{-0.5cm}
\end{figure}

\begin{figure*}[!th]
\vspace{-0.1cm}
	\centering
	\begin{subfigure}{0.32\textwidth}
		\centering
		\includegraphics[scale=0.42]{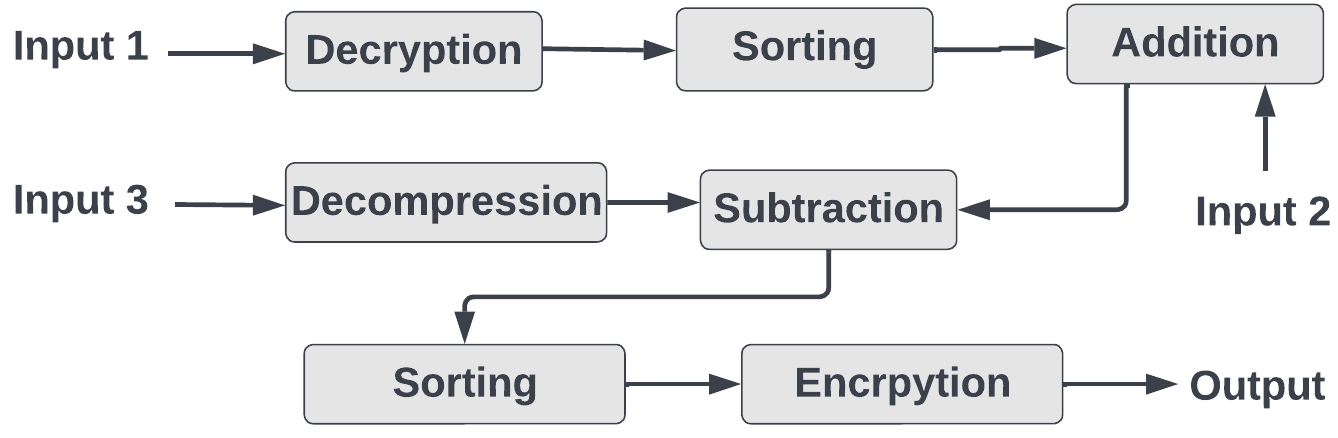}
		\caption{An example graph of the synthetic dataset.}
		\label{Figure:mixing_syn_acc}
	\end{subfigure} \hfil
	\begin{subfigure}{0.32\textwidth}
	    \centering
	    \includegraphics[scale=0.42]{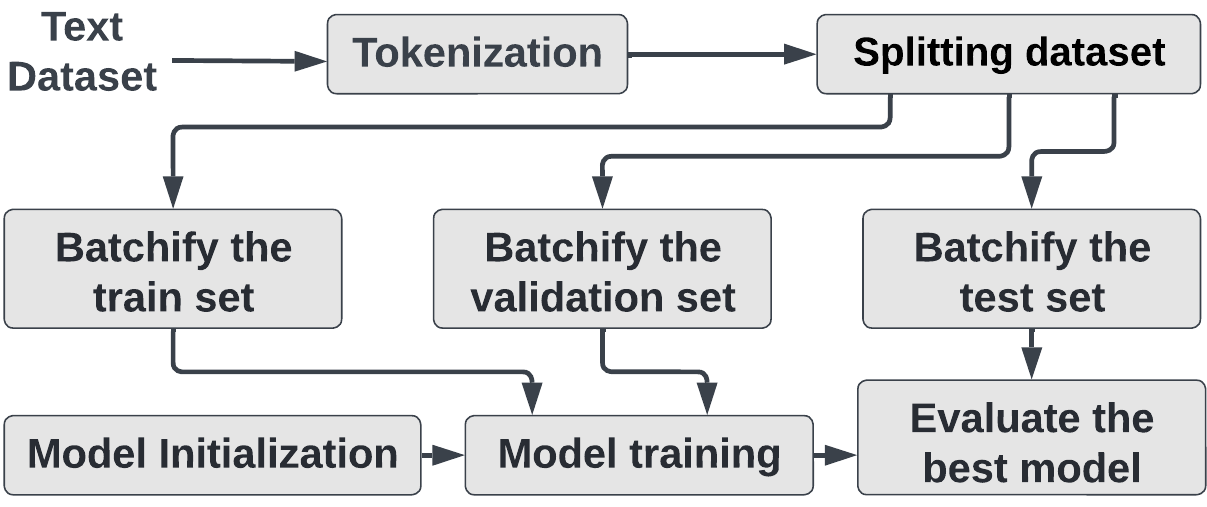}
	    \caption{An example graph of the ML/AI dataset.}
	    \label{Figure:mixing_ml_acc}
	\end{subfigure}\hfil
	\begin{subfigure}{0.32\textwidth}
	    \centering
	    \includegraphics[scale=0.42]{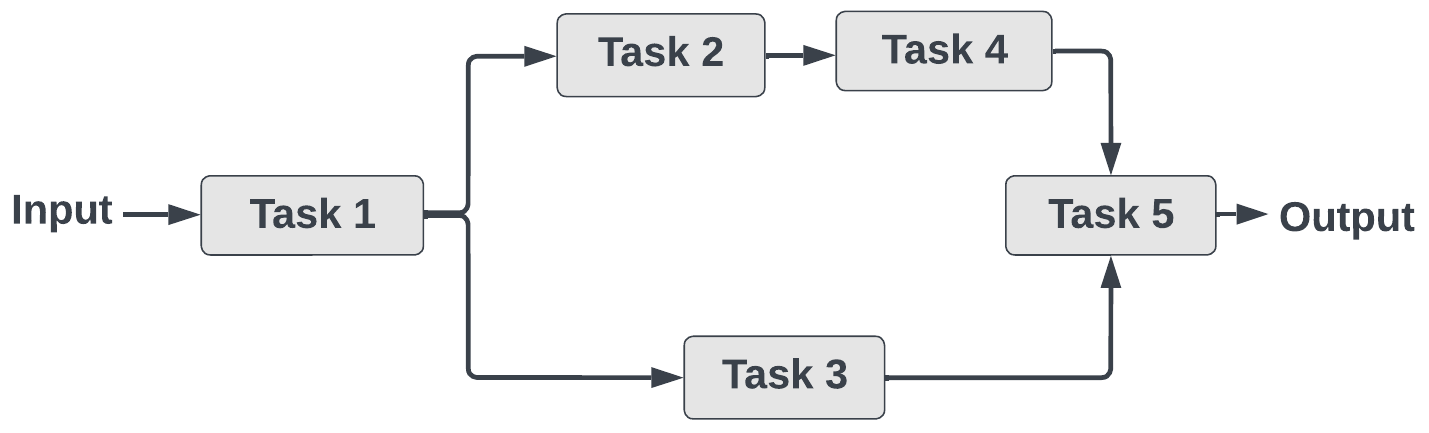}
	    \caption{An example graph of the Alibaba  dataset.}
	    \label{Figure:mixing_alibaba_acc}
	\end{subfigure}\hfil
	\vspace{-0.1cm}
	\caption{Representative examples of computation graphs from the datasets used for the evalution of \sol.}
	\label{figure:example_graphs}
	\vspace{-0.4cm}
\end{figure*}

To evaluate the \sol prototype, we use a Decision Tree \cite{quinlan1987simplifying}, a k-Nearest Neighbors (kNN) algorithm \cite{altman1992introduction}, and a Neural Network (NN) \cite{dayhoff1990neural} as the attack models that identify executed computations. Figure \ref{Figure:eval_steps} shows the process we followed for our evaluation. We utilize three computation datasets to train the attack models: a synthetic dataset, an ML/AI dataset, and a real-world dataset from Alibaba cloud \cite{alibaba}. We summarize the characteristics of these datasets in Table~\ref{Table:datasets}, we present representative examples of computation graphs for each dataset in Figure~\ref{figure:example_graphs}, and we discuss them further below. 

Once an attack model is trained, we measure and use the accuracy of the trained model as our baseline (\ie case of no anonymization) to quantify the effectiveness of \sol. Subsequently, \sol (either the \solOne, the \mix, or the \hybrid mode) is applied to a computation to anonymize it, the anonymized computation graph is executed, and its features are collected. The collected features are fed to the trained attack model to predict the identity of the computation and we again measure the accuracy of the model for anonymized computations. We eventually compare the accuracy of the attack model measured before and after anonymization. The reduction in accuracy indicates the effectiveness of \sol. We also measure the time required by \sol to convert an original graph to the corresponding anonymized graph and the overhead (in terms of time, and CPU and memory usage) introduced to the execution process because of anonymization. Finally, to evaluate \sol under adaptive attack models (as we discussed in Section~\ref{sec:threat}), we train attack models with both original (without anonymization) and anonymized computation graph data and follow the process described above to measure the models' accuracy.


\begin{table*}[t]
\centering
\caption{Datasets used for the evaluation of \sol and their characteristics.}
\vspace{-0.2cm}
\resizebox{0.8\textwidth}{!}{%
\label{Table:datasets}
\begin{tabular}{|c|c|c|c|c|c|c|}
\hline
\textbf{Dataset}      & \textbf{Features}                                                               & \textbf{Nature}             & \textbf{\begin{tabular}[c]{@{}c@{}}Average Number \\ of Nodes\end{tabular}} & \textbf{\begin{tabular}[c]{@{}c@{}}Average Node \\ Degree\end{tabular}} & \textbf{\begin{tabular}[c]{@{}c@{}}Total Number \\ of Samples\end{tabular}} & \textbf{\begin{tabular}[c]{@{}c@{}}Total Number \\ of Classes\end{tabular}} \\ \hline
\begin{tabular}[c]{@{}c@{}}Synthetic \\ dataset \end{tabular} & \begin{tabular}[c]{@{}c@{}}CPU usage, memory\\ usage, completion time,\\ inputs/outputs\end{tabular} & \begin{tabular}[c]{@{}c@{}} Popular  algorithmic operations (sorting, \\ searching, hashing, encryption, \\ decryption, compression, decompression). \end{tabular} & 4.2 $\pm$ 2.1  & 3.0 $\pm$ 0.9 & 1000 & 10\\ \hline

\begin{tabular}[c]{@{}c@{}} ML/AI graph \\ dataset \end{tabular}   & \begin{tabular}[c]{@{}c@{}} CPU usage, memory usage, \\ inputs/outputs, \\ completion time \end{tabular}    & \begin{tabular}[c]{@{}c@{}} Image processing, image and text \\classification, language translation,\\ automatic image annotation \end{tabular}  & 6.4 $\pm$ 2.4 & 2.2 $\pm$ 0.8 & 1000 & 10\\ \hline

\begin{tabular}[c]{@{}c@{}} Alibaba Cloud \\ dataset \end{tabular} &\begin{tabular}[c]{@{}c@{}} CPU usage, memory \\ usage, completion times \end{tabular}                                                                                  & Data-center cluster dataset & 2.7 $\pm$ 1.4                                                                             &     2.0 $\pm$ 0.6 & 20000 & 10\\ \hline

\end{tabular}

}

\vspace{-0.2cm}
\end{table*}

\subsubsection{Datasets}


\noindent\textbf{Synthetic dataset:} 
This dataset includes 1000 samples of 10 different computation graphs, which are formed by interconnecting nodes that represent popular algorithmic operations (sorting, searching, hashing, encryption, decryption, compression, and decompression). We divided the computations of this dataset into two categories: delay-sensitive and delay-tolerant computations. For our evaluation, we considered delay-sensitive computations as more important than delay-tolerant computations (\ie delay-sensitive computations need to be executed as quickly as possible). Each sample contains the number of inputs and outputs, the input and output sizes, the completion time of each computation, and CPU and memory usage information for each computation.


\noindent\textbf{ML/AI graph dataset:} This dataset includes 1000 samples of 10 different ``off-the-shelf" real-world machine learning and artificial intelligence graphs. Each graph includes the workflow to train and validate a machine learning model. Different steps, such as pre-processing a dataset, building a model architecture, training the model with the dataset, and evaluating the model performance, are represented as nodes of each graph. We executed these graphs with different hyper-parameters and training dataset sizes to collect features for these workloads. We considered this dataset for our evaluation, since a wide range of real-world use cases and services are related to ML/AI and are offered by commercial cloud providers, such as the Amazon SageMaker, the Azure Applied AI Services, and Vertex AI by Google Cloud. Each sample contains the number of inputs and outputs, the input and output sizes, the completion time of each computation, and CPU and memory usage information for each computation.


\noindent\textbf{Alibaba cloud dataset:} This dataset contains 20000 samples of 10 batch jobs collected from 4000 machines over a period of eight days. Although the dataset comes with information on servers, containers, and batch jobs, we utilize the data related to batch jobs to evaluate \sol, since our main goal is to identify the executed computations. A job (\ie a computation graph) is a set of tasks (\ie nodes) that are dependent on each other and form a DAG. The jobs are categorized into multiple types that we intend to identify with our attack models.
Each sample includes CPU and memory usage information in a normalized format, as well as completion times.


\subsubsection{Evaluation Metrics}


We use the following metrics to evaluate \sol: (i) the attack model's identification accuracy; and (ii) the overhead introduced due to anonymization. The reduction of the attack model's accuracy with respect to the baseline model (\ie no anonymization) indicates the effectiveness of \sol. We quantify overhead as the additional time and computing resources (CPU and memory usage) needed to complete an anonymized computation as compared to the original computation with no anonymization.


\subsection {Evaluation Results}
\label{subsec:results}


\subsubsection{\solOne mode} 

\begin{figure*}[!th]
	\centering
	\begin{subfigure}{0.33\textwidth}
		\centering
		\includegraphics[scale=0.2]{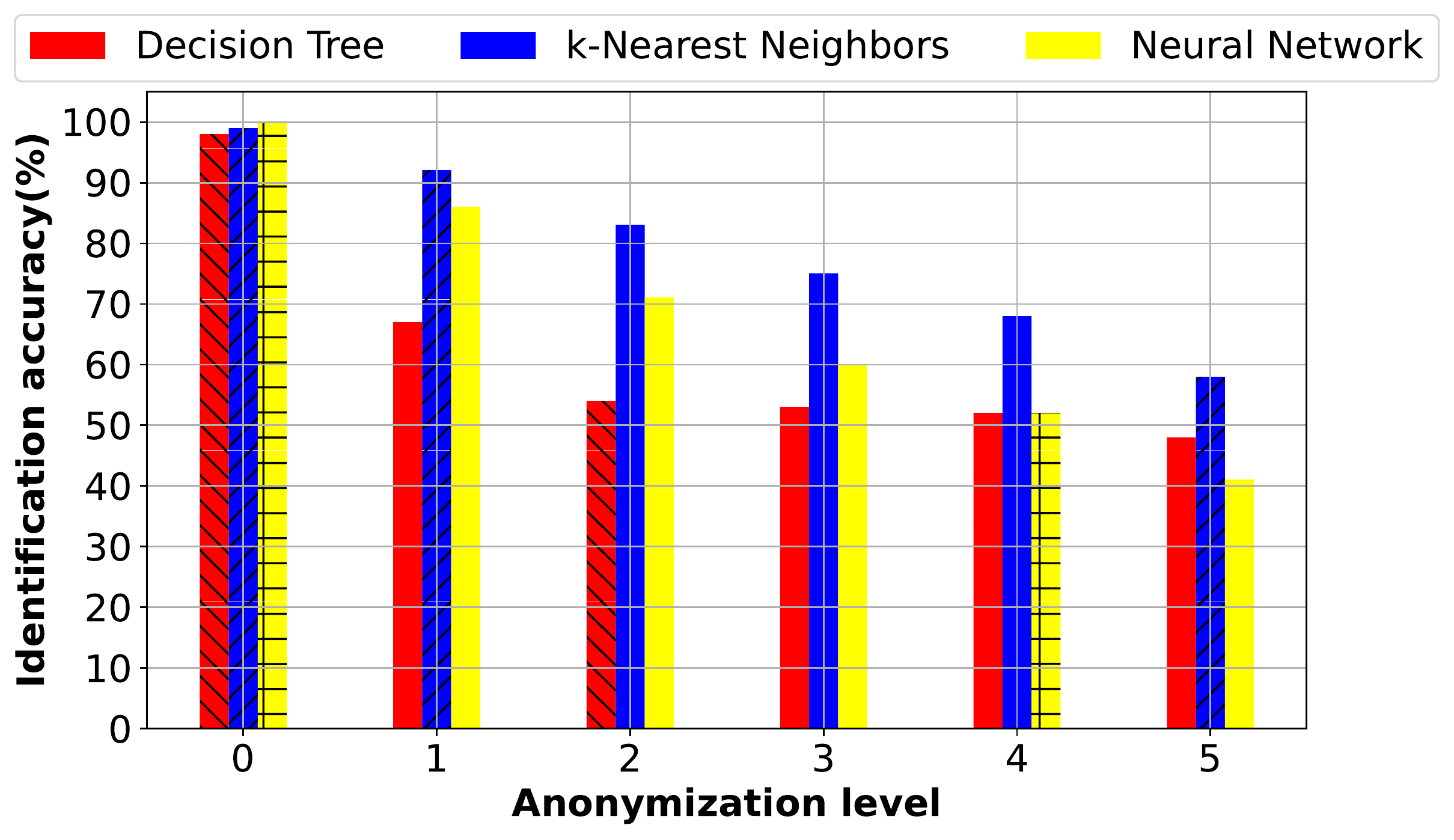}
         \vspace{-0.15cm}
		\caption{Synthetic dataset.}
		\label{Figure:solOne_syn_acc_level}
	\end{subfigure} \hfil
	\begin{subfigure}{0.33\textwidth}
	    \centering
	    \includegraphics[scale=0.2]{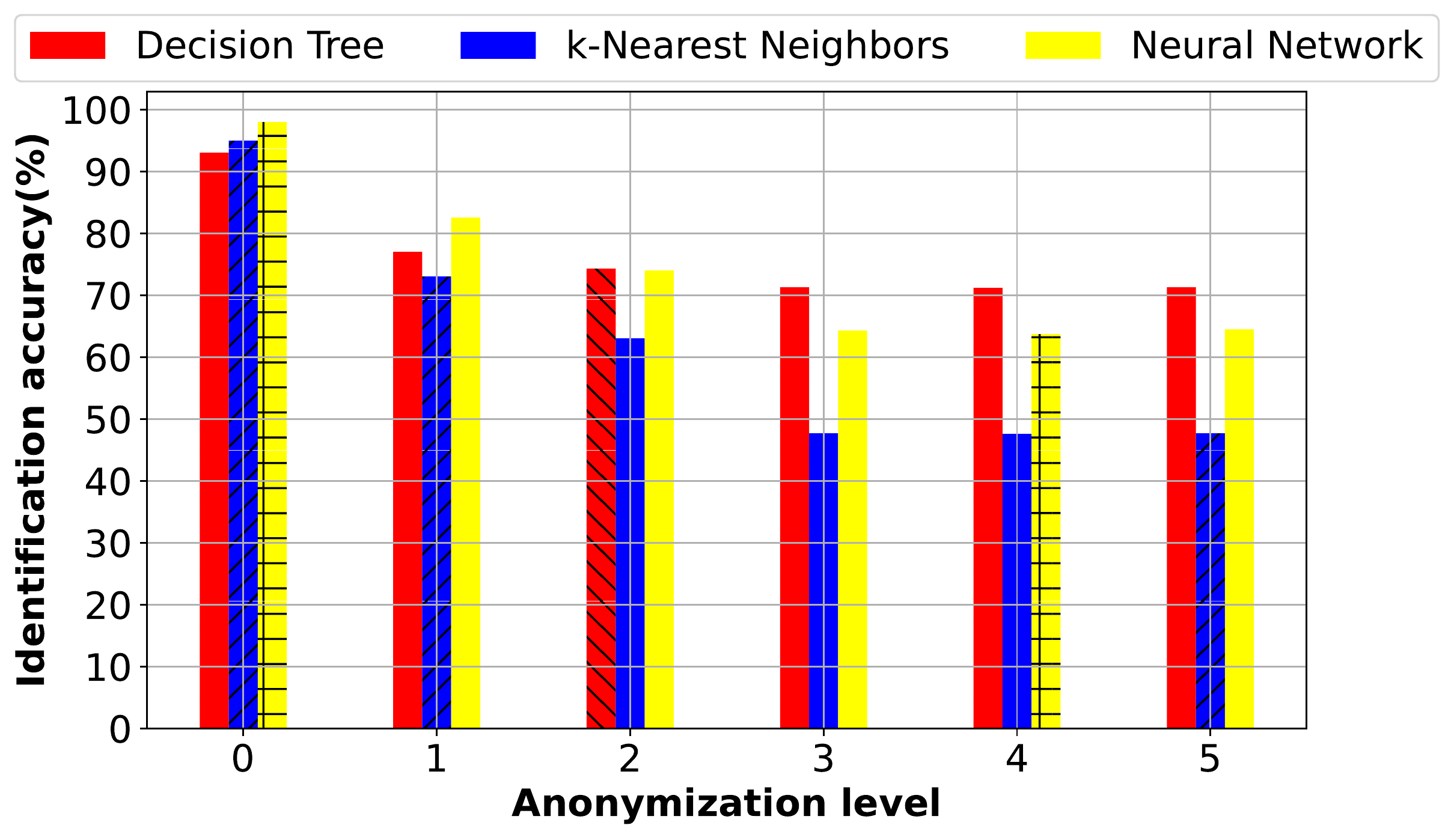}
         \vspace{-0.15cm}
	    \caption{ML/AI dataset}
	    \label{Figure:solOne_ml_acc_level}
	\end{subfigure}\hfil
	\begin{subfigure}{0.33\textwidth}
	    \centering
	    \includegraphics[scale=0.2]{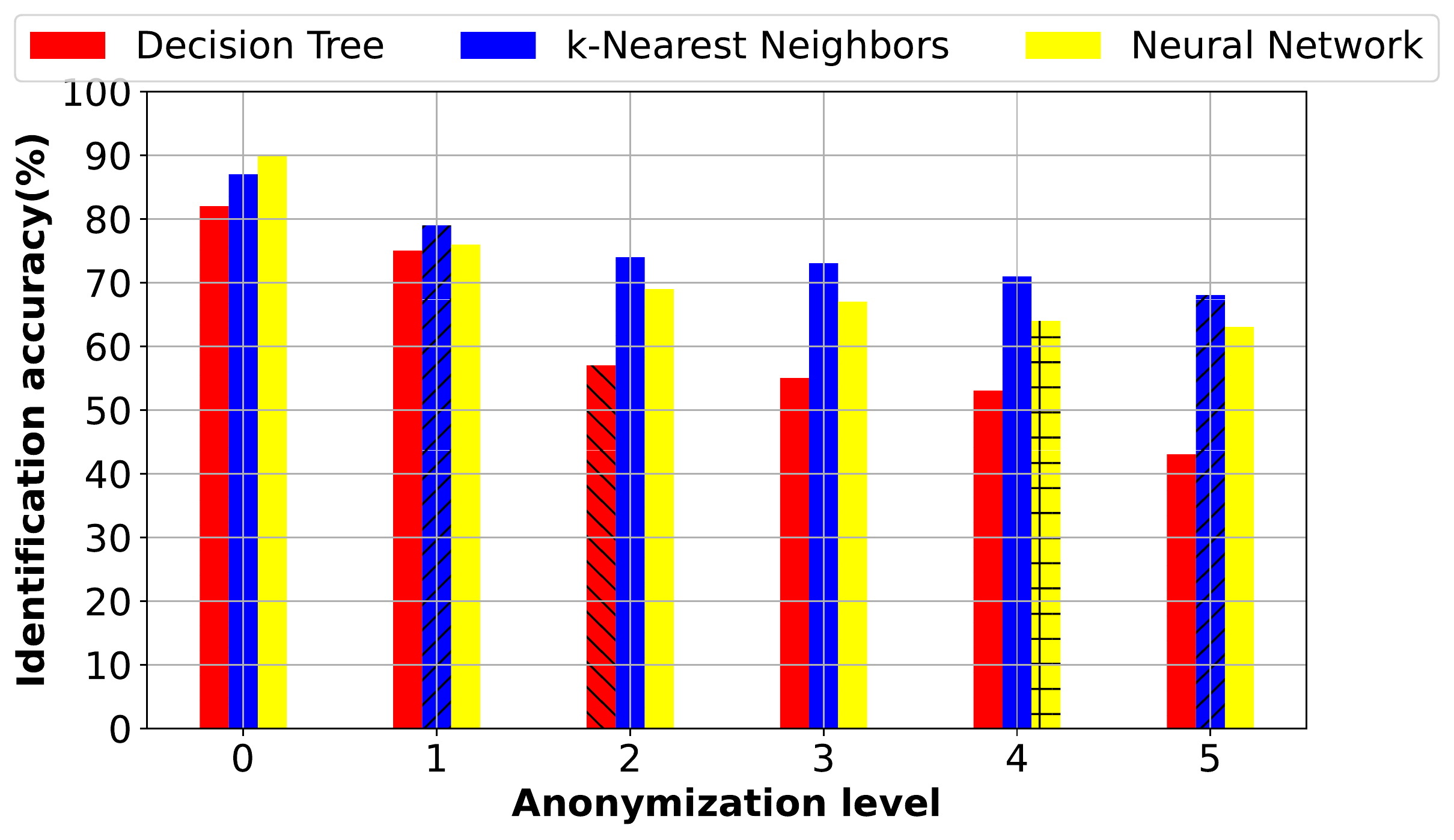}
         \vspace{-0.15cm}
	    \caption{Alibaba cloud dataset.}
	    \label{Figure:alibaba_syn_acc_level}
	\end{subfigure}\hfil
	\vspace{-0.2cm}
	\caption{Identification accuracy for different attack models, datasets, and anonymization levels using the \solOne mode.}
	\label{figure:solOne_acc_level}
	\vspace{-0.7cm}
\end{figure*}

\noindent\textbf{Effect of anonymization on accuracy:} In Figure \ref{figure:solOne_acc_level}, we present the computation identification accuracy results for different attack models when computations are anonymized through the \solOne mode. For all three datasets, each attack model demonstrates a similar trend. Specifically, the computation identification accuracy of all three models decreases as the anonymization level increases. \solOne is able to achieve up to a 60\% accuracy reduction. However, as we increase the anonymization level, the effectiveness of \solOne (\ie the reduction of accuracy) becomes smaller, since the \solOne mode adds sufficient randomness to the original computation graph. 
The results also show that the models have relatively better accuracy for the synthetic and the ML/AI datasets for all anonymization levels as compared to the Alibaba cloud dataset. This is due to the fact that the Alibaba cloud dataset contains fewer features (\eg no input and output data sizes), which can be used for the identification of computations compared to the other datasets.

\begin{figure}[!t]
\vspace{-0.1cm}
 \centering
 \includegraphics[scale=0.25]{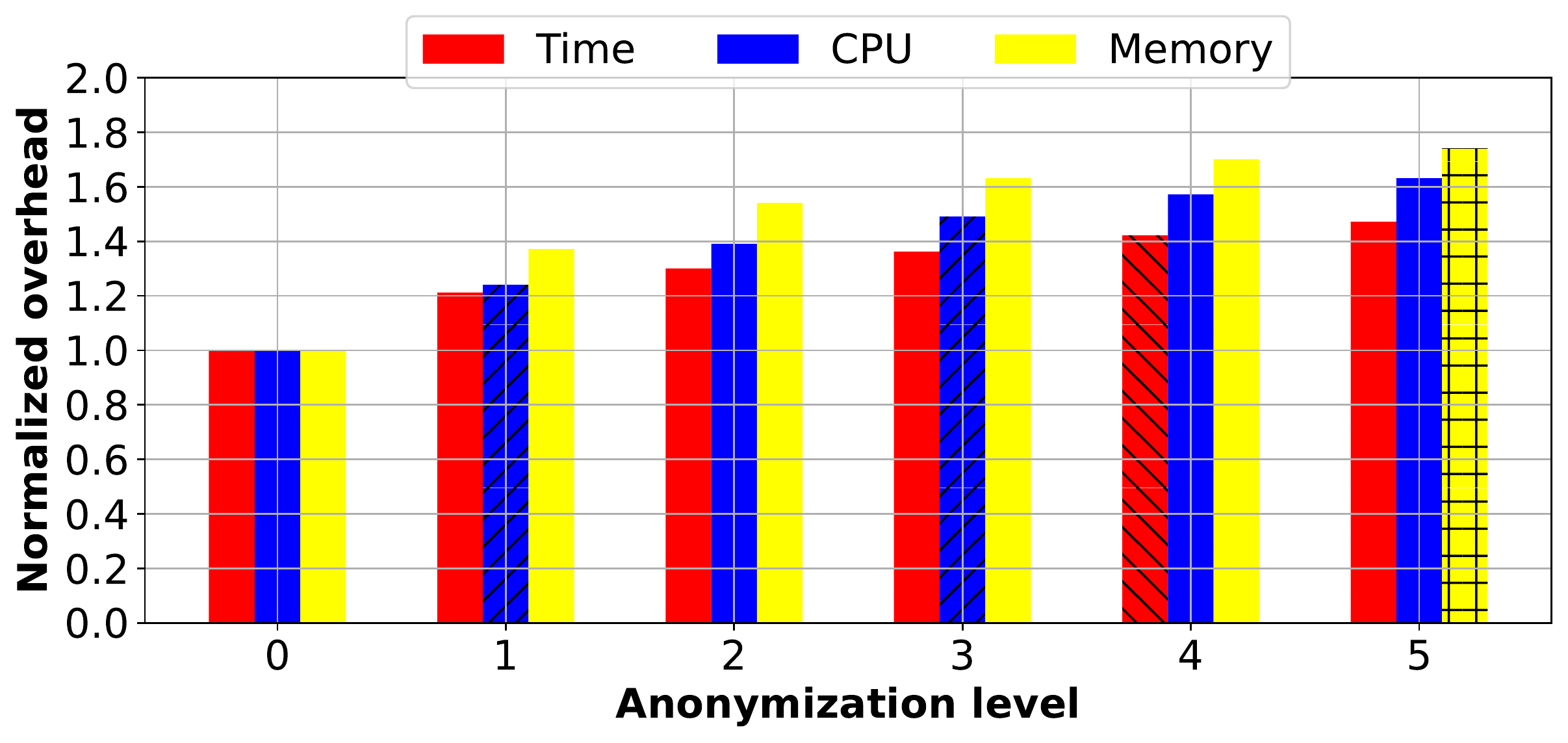}
 \vspace{-0.3cm}
 \caption{Normalized overheads of the \solOne mode for different anonymization levels} 
 \label{Figure:solOne_overhead_level}
 \vspace{-0.7cm}
\end{figure}

\noindent\textbf{Effect of anonymization on overhead:} In Figure \ref{Figure:solOne_overhead_level}, we present the time, CPU usage, and memory usage overhead for the execution of computations anonymized through the \solOne mode for different anonymization levels in a normalized form. Specifically, we normalize the required time to execute each anonymized computation by the required time to complete the same computation without anonymization. In the same manner, we also normalize the CPU and memory usage overheads. As Figure \ref{Figure:solOne_overhead_level} indicates, all three types of overhead show the same trend. The overhead to execute a computation increases as the anonymization level increases. This is due to the fact that a higher level of anonymization increases the complexity of the overall anonymization process (\eg  more ``fake'' nodes are added to computation graphs, padding of inputs and outputs becomes more extensive). Note that the results of Figure~\ref{Figure:solOne_overhead_level} hold for all used datasets. 


\noindent\textbf{Impact of anonymizing different features:} In Table \ref{tab:impact}, we present results on how anonymizing different computation features can impede the attack model's ability to identify computations. Our results demonstrate that adding padding only to inputs or outputs or even both is unable to reduce the attack model's effectiveness in a substantial manner. Nevertheless, anonymizing the completion time of computations by adding ``fake'' nodes reduces the accuracy of the attack model substantially as the anonymization level increases. Combining input and output padding and completion time anonymization results in the highest reduction of the attack model's accuracy.


\begin{table*}%
\centering
\caption{Impact of anonymizing different features on attack model accuracy (\%)}
\vspace{-0.2cm}
\resizebox{0.65\textwidth}{!}{%
\label{tab:impact}
\centering
\begin{tabular}{|c|c|c|c|c|c|}
\hline
\multicolumn{1}{|l|}{\textbf{\begin{tabular}[c]{@{}c@{}}Anonymization \\ level\end{tabular}}} & \textbf{\begin{tabular}[c]{@{}c@{}}Only input\\ padding\end{tabular}} &
\textbf{\begin{tabular}[c]{@{}c@{}}Only output\\ padding\end{tabular}} & \textbf{\begin{tabular}[c]{@{}c@{}}Both input and\\ output padding\end{tabular}} & 
\textbf{\begin{tabular}[c]{@{}c@{}}Only completion\\ time anonymization\end{tabular}} & \textbf{\begin{tabular}[c]{@{}c@{}}All features\\ anonymization\end{tabular}} \\ \hline
1 & 99.1 & 99.1 & 99.0 & 95.0 & 91.7                       \\ \hline
2 & 99.0 & 98.6 & 98.8 & 87.0 & 83.3   \\ \hline
3 & 99.1 & 97.5 & 98.5 & 77.0 & 75.4  \\ \hline
4 & 98.9 & 94.5 & 96.2 & 73.0 & 68.0   \\ \hline
5 & 99.1 & 91.2 & 95.2 & 67.0 & 57.9  \\ \hline

\end{tabular}
}
\vspace{-0.4cm}
\end{table*}

\begin{table}%
\vspace{-0.2cm}
\centering
\caption{Required time to anonymize a computation graph using the \solOne mode}
\vspace{-0.2cm}
\resizebox{0.65\columnwidth}{!}{%
\label{tab:solOne_conversation_time}
\centering
\begin{tabular}{|c|c|}
\hline
\multicolumn{1}{|l|}{\textbf{\begin{tabular}[c]{@{}c@{}}Anonymization level\end{tabular}}} & \textbf{\begin{tabular}[c]{@{}c@{}}Required time to anonymize (ms) \end{tabular}}  \\ \hline
1 &  7  \\ \hline
2 &  14  \\ \hline
3 &  25 \\ \hline
4 &  34  \\ \hline
5 &  40  \\ \hline

\end{tabular}
}
\vspace{-0.4cm}
\end{table}

\noindent\textbf{Required time to anonymize a computation graph:} Table \ref{tab:solOne_conversation_time} shows the required time to anonymize a graph on the fly using the \solOne mode (step 3a in Figure \ref{Figure:workflow} of our design) for different anonymization levels. Our results indicate that the required time increases as the anonymization level increases. For greater anonymization levels, the \solOne mode needs to introduce more ``randomness'' to computations which, subsequently, increases the anonymization time. Overall, our results show that the \solOne mode needs less than 40ms to anonymize a computation. In cases of delay-sensitive and highly private computations (which may require a greater anonymization level), computations can be anonymized in advance (pre-anonymized) as we discussed in Section~\ref{subsec:solone}. This amortizes the overall cost and eliminates the need to conduct computation anonymization on the fly.

\subsubsection{\mix mode}
\begin{figure*}[!th]
\vspace{-0.1cm}
	\centering
	\begin{subfigure}{0.3\textwidth}
		\centering
		\includegraphics[scale=0.2]{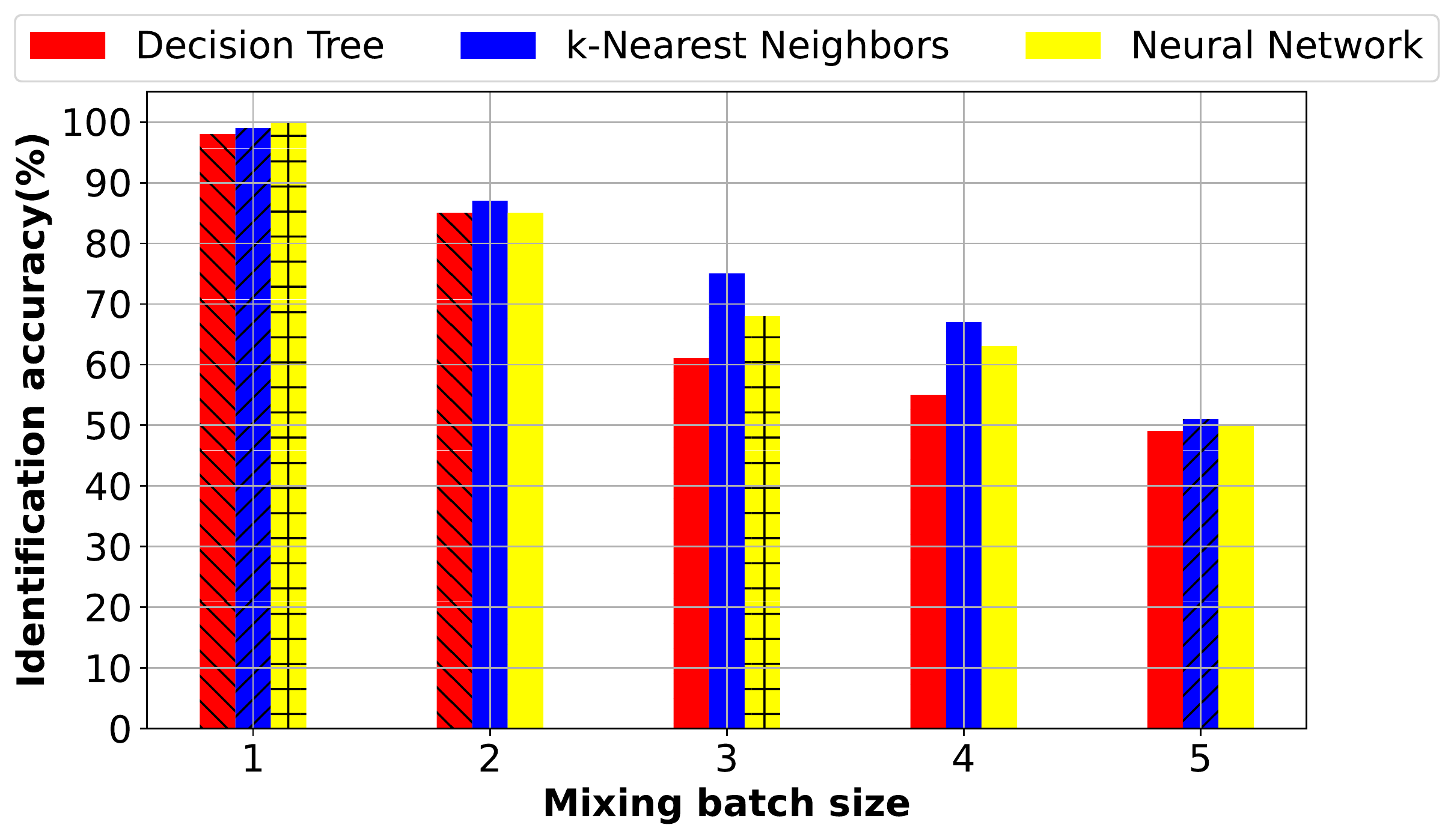}
         \vspace{-0.15cm}
		\caption{Synthetic dataset.}
		\label{Figure:mixing_syn_acc}
	\end{subfigure} \hfil
	\begin{subfigure}{0.3\textwidth}
	    \centering
	    \includegraphics[scale=0.2]{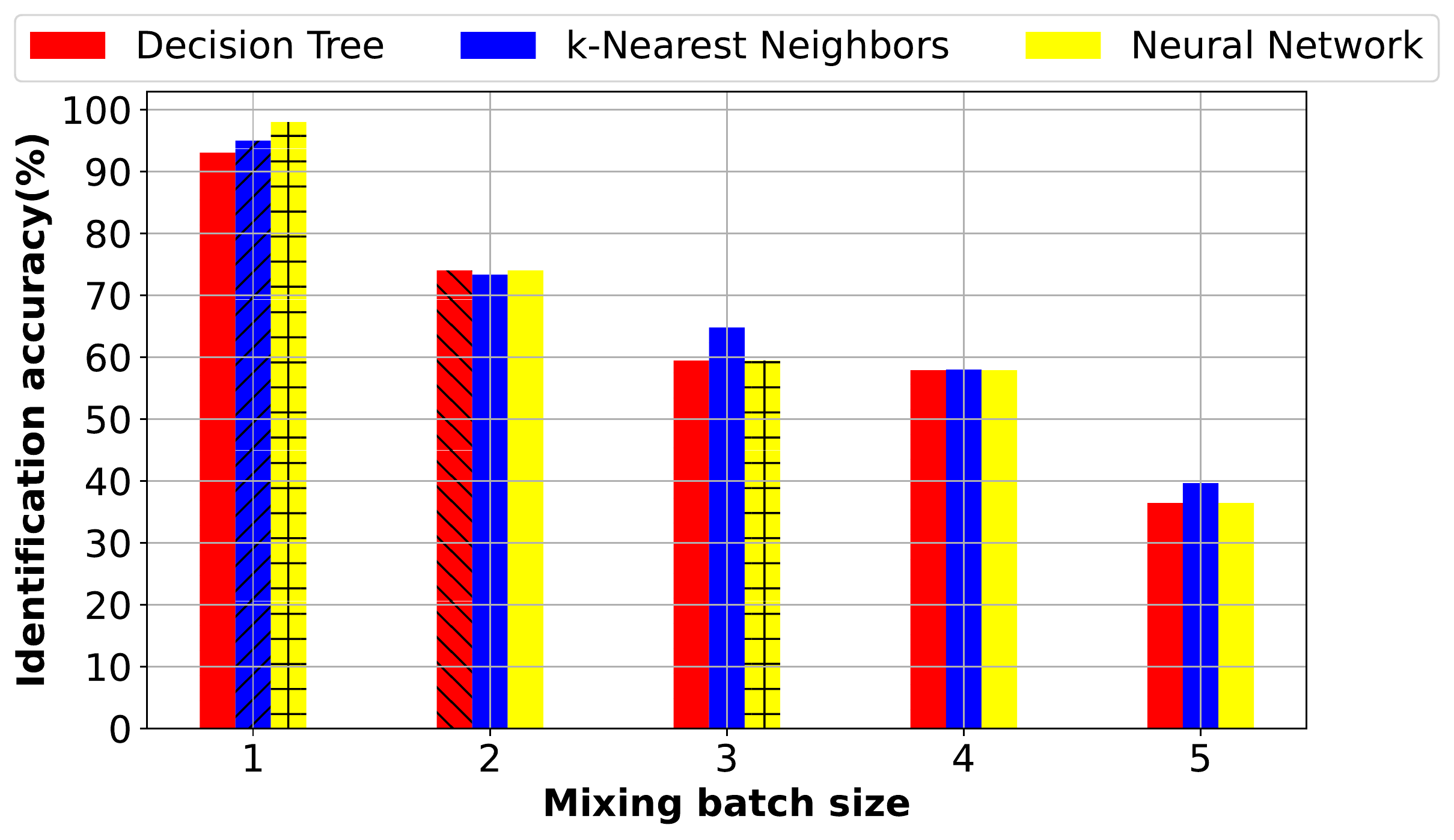}
         \vspace{-0.15cm}
	    \caption{ML/AI dataset}
	    \label{Figure:mixing_ml_acc}
	\end{subfigure}\hfil
	\begin{subfigure}{0.3\textwidth}
	    \centering
	    \includegraphics[scale=0.2]{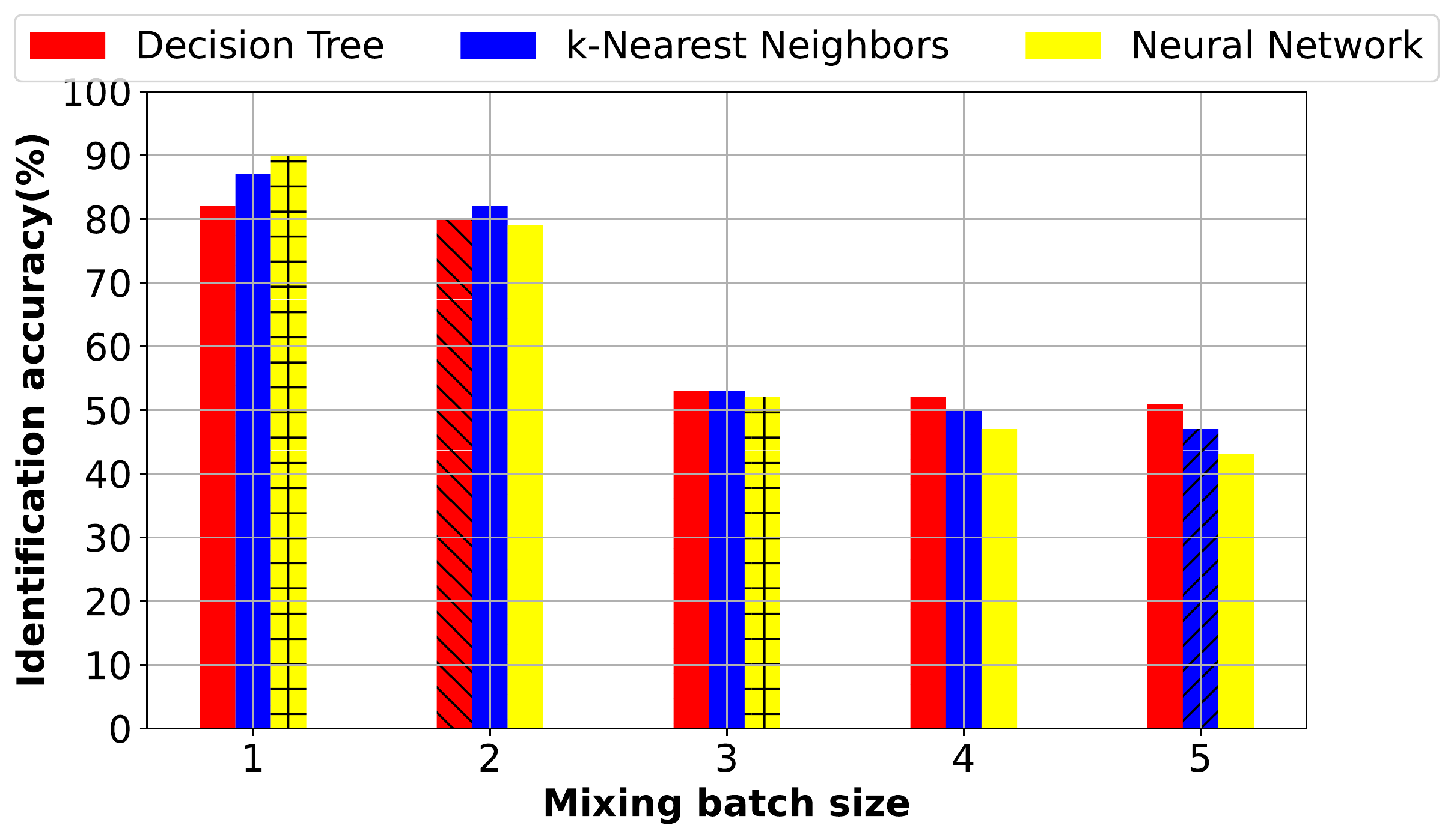}
         \vspace{-0.15cm}
	    \caption{Alibaba cloud dataset.}
	    \label{Figure:mixing_alibaba_acc}
	\end{subfigure}\hfil
	\vspace{-0.2cm}
	\caption{Identification accuracy for different attack models, datasets, and batch sizes using the sequential \mix mode.}
	\label{figure:mixing_acc_batch}
	\vspace{-0.4cm}
\end{figure*}

\begin{figure*}[!th]
\vspace{-0.1cm}
	\centering
	\begin{subfigure}{0.3\textwidth}
		\centering
		\includegraphics[scale=0.2]{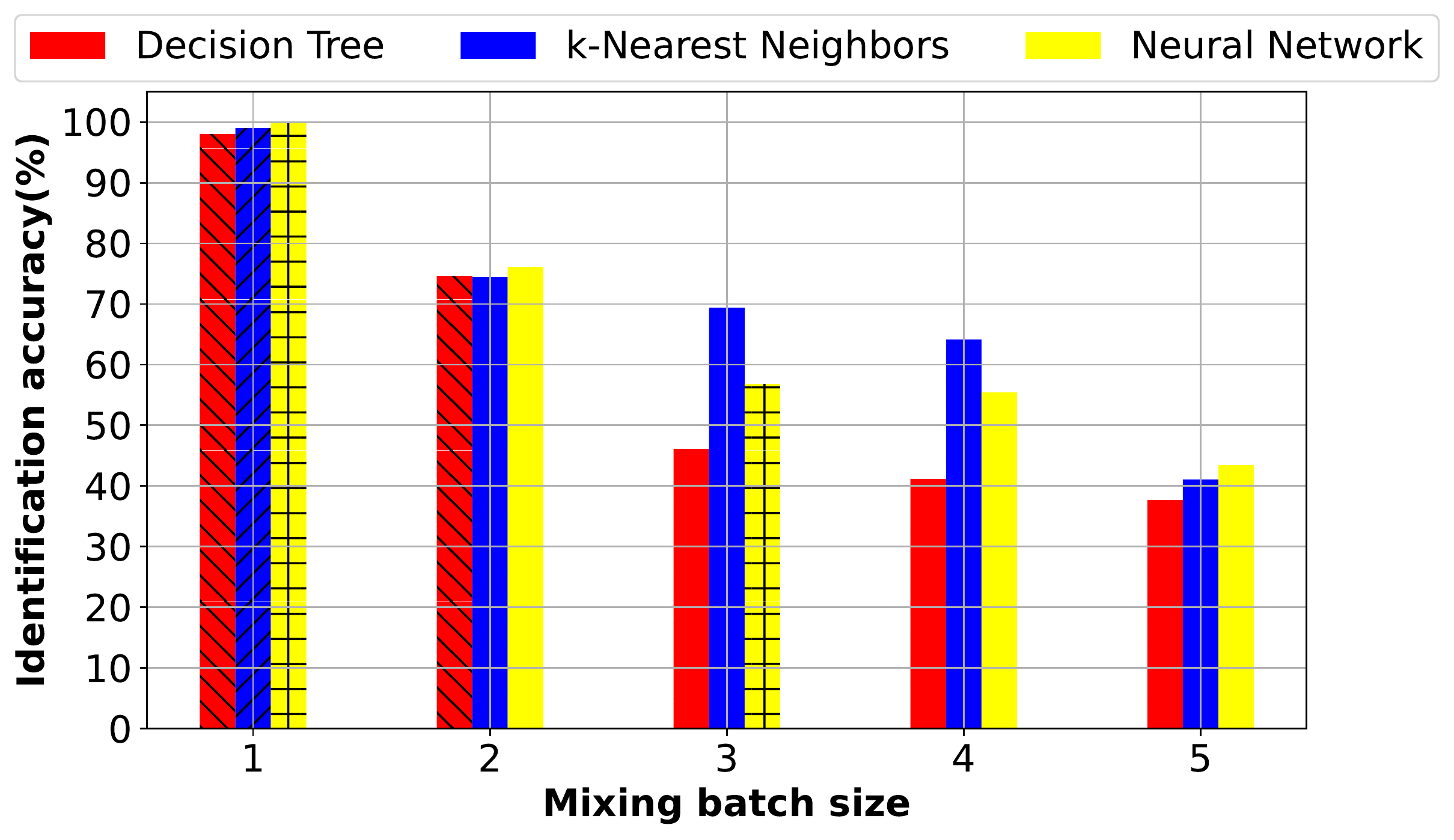}
         \vspace{-0.15cm}
		\caption{Synthetic dataset.}
		\label{Figure:mixing_parallel_syn_acc}
	\end{subfigure} \hfil
	\begin{subfigure}{0.3\textwidth}
	    \centering
	    \includegraphics[scale=0.2]{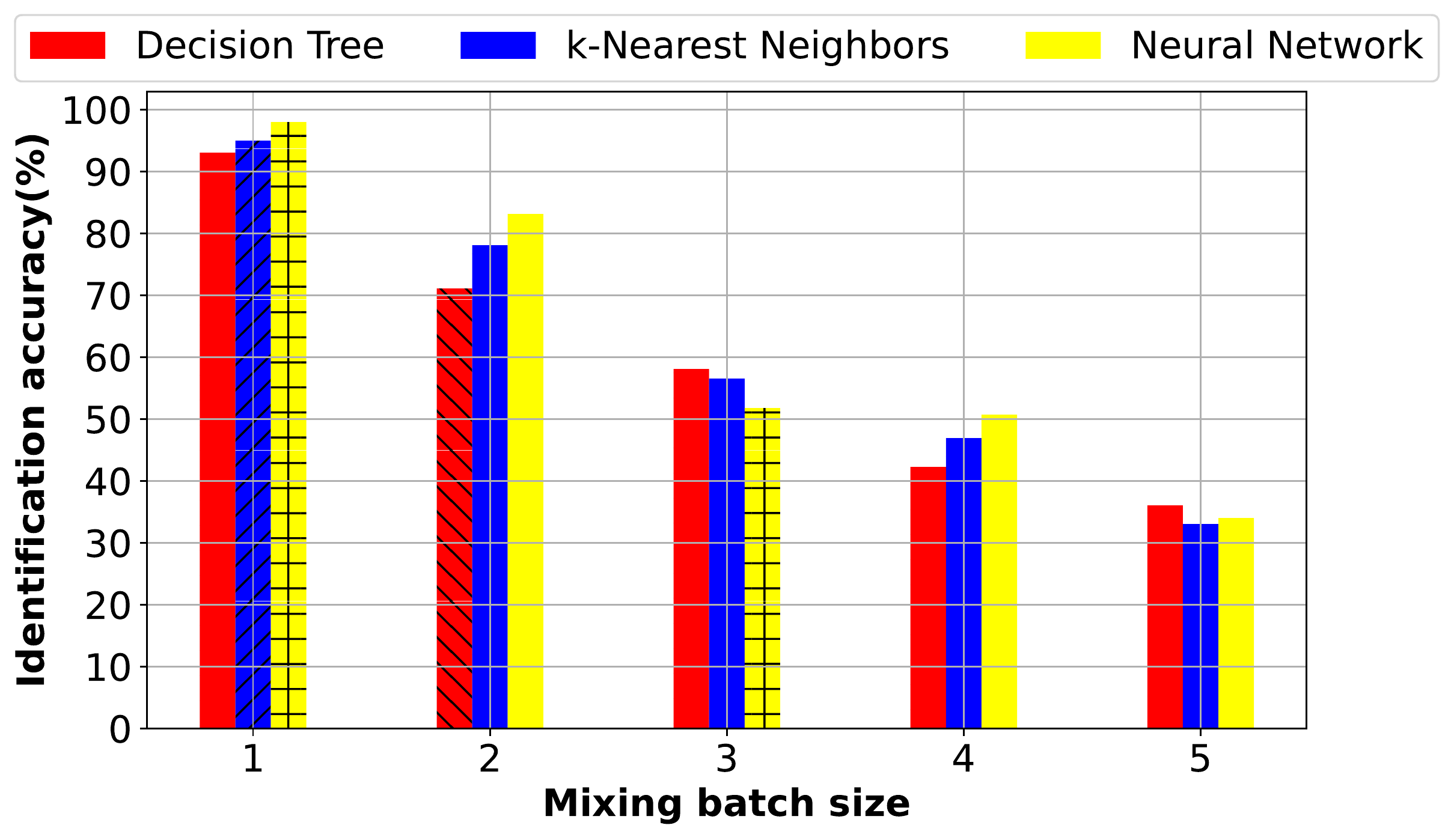}
         \vspace{-0.15cm}
	    \caption{ML/AI dataset}
	    \label{Figure:mixing_parallel_ml_acc}
	\end{subfigure}\hfil
	\begin{subfigure}{0.3\textwidth}
	    \centering
	    \includegraphics[scale=0.2]{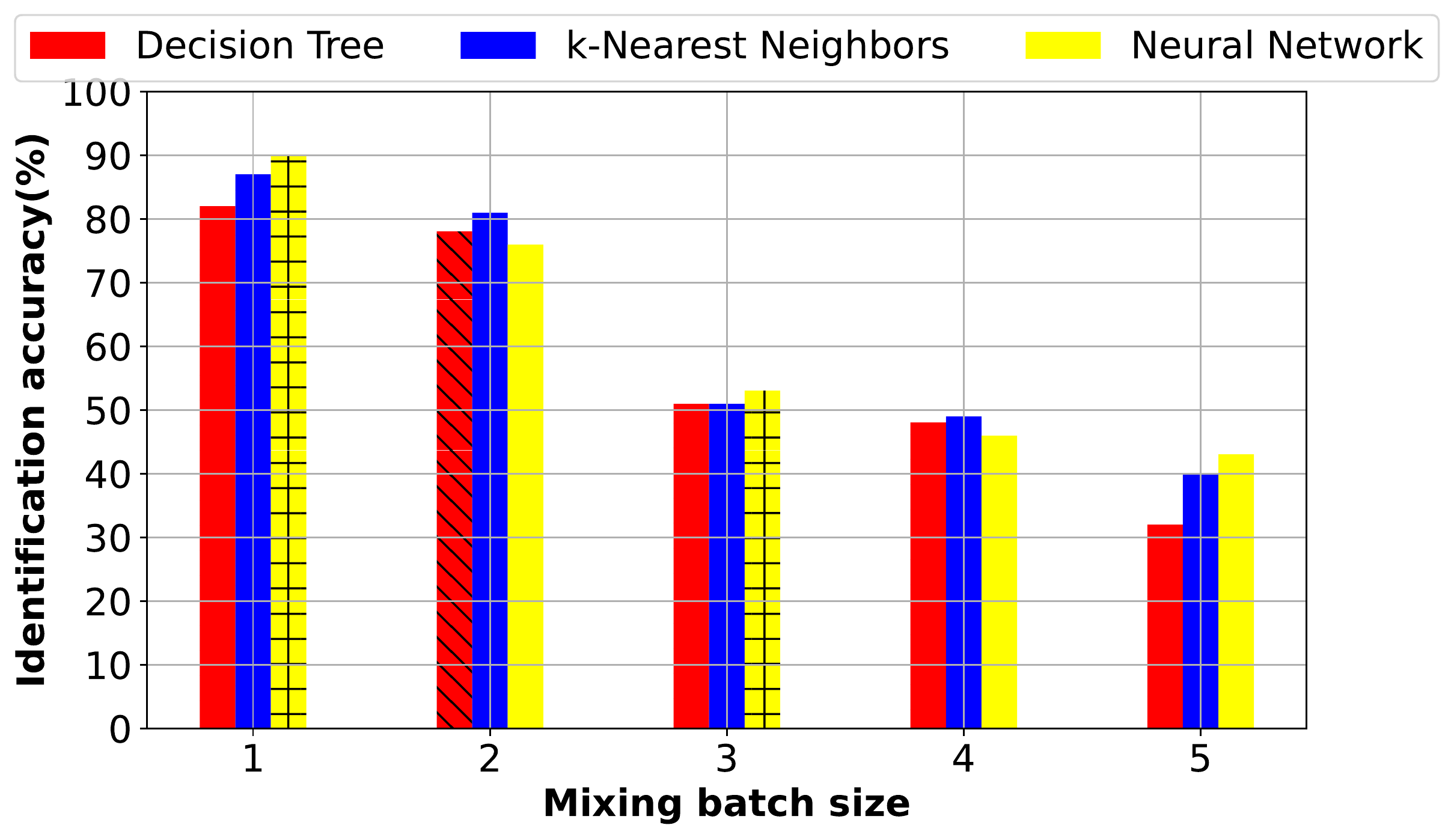}
         \vspace{-0.15cm}
	    \caption{Alibaba cloud dataset.}
	    \label{Figure:mixing_parallel_alibaba_acc}
	\end{subfigure}\hfil
	\vspace{-0.2cm}
	\caption{Identification accuracy for different attack models, datasets, and batch sizes using the parallel \mix mode.}
	\label{figure:mixing_parallel_acc_batch}
	\vspace{-0.5cm}
\end{figure*}

\noindent\textbf{Effect of anonymization on accuracy:} Figures \ref{figure:mixing_acc_batch} and \ref{figure:mixing_parallel_acc_batch} show how the accuracy of attack models is impacted as the number of computations shuffled together (mixing batch size) varies for the sequential and the parallel \mix modes respectively. As the batch size increases, the accuracy of all models for all three datasets decreases for both the sequential and parallel modes, since a greater batch size results in mixing more computations. This makes it harder for attackers to identify the executed computations.

In most cases, the parallel \mix mode impairs the ability to identify computations slightly more as compared to the sequential \mix mode. In the parallel mode, the completion time of a computation graph can be either lower or higher than the completion time of the same graph without anonymization, since there are multiple available threads and the nodes of the graph can be executed by one or more threads at the same time. On the other hand, in the sequential mode, the completion time of a graph is always higher than the completion time of the same graph without anonymization. Due to this randomness introduced in the completion time, the parallel \mix mode is relatively more effective than the sequential \mix mode. Overall, with the \mix mode, we are able to reduce the identification accuracy of attack models by up to 50\%.

\begin{figure}[!t]
\vspace{-0.3cm}
 \centering
 \includegraphics[scale=0.25]{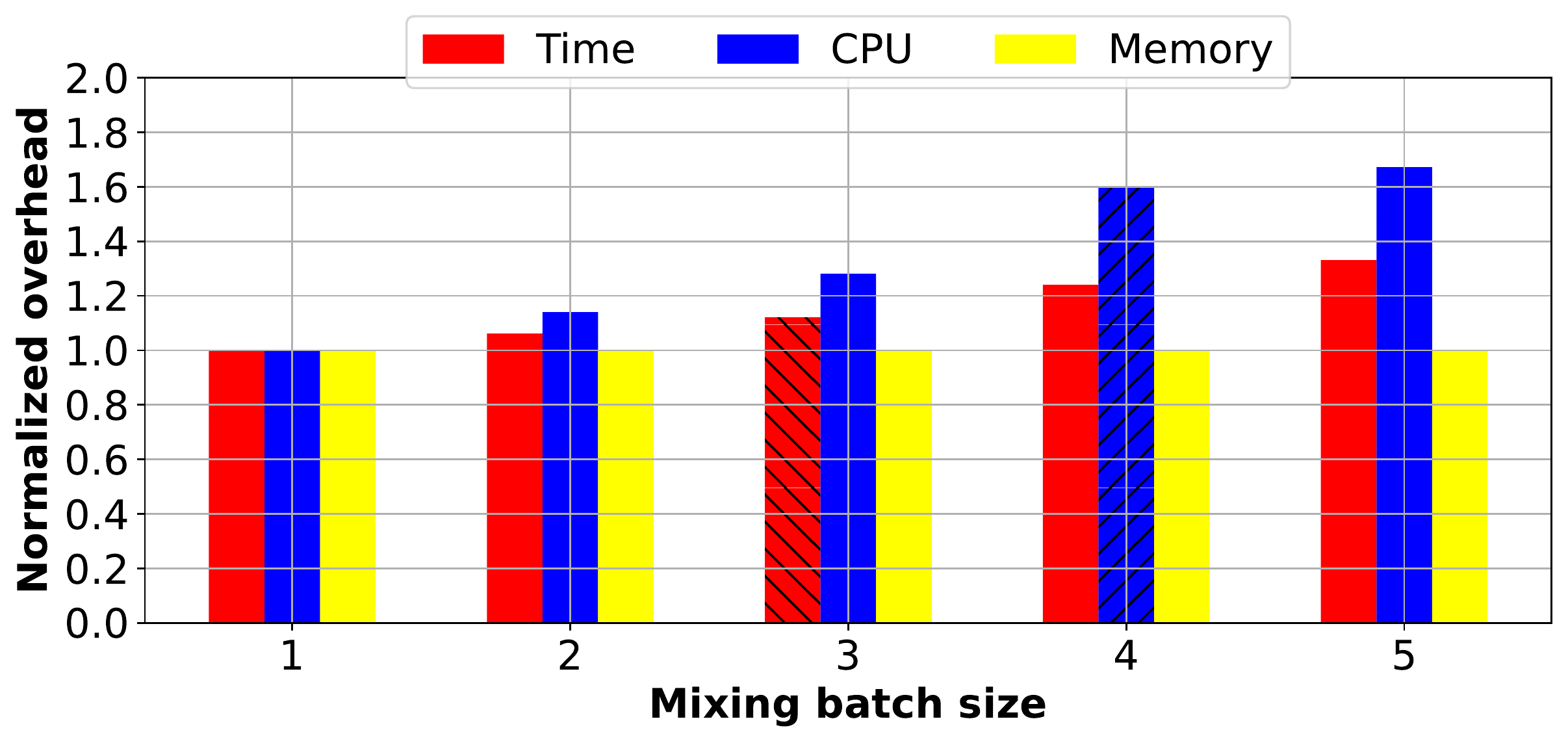}
 \vspace{-0.2cm}
 \caption{Normalized overheads of the sequential \mix mode for different batch sizes} 
 \label{Figure:mixing_overhead}
 \vspace{-0.3cm}
\end{figure}

\begin{figure}[!t]
\vspace{-0.3cm}
 \centering
 \includegraphics[scale=0.25]{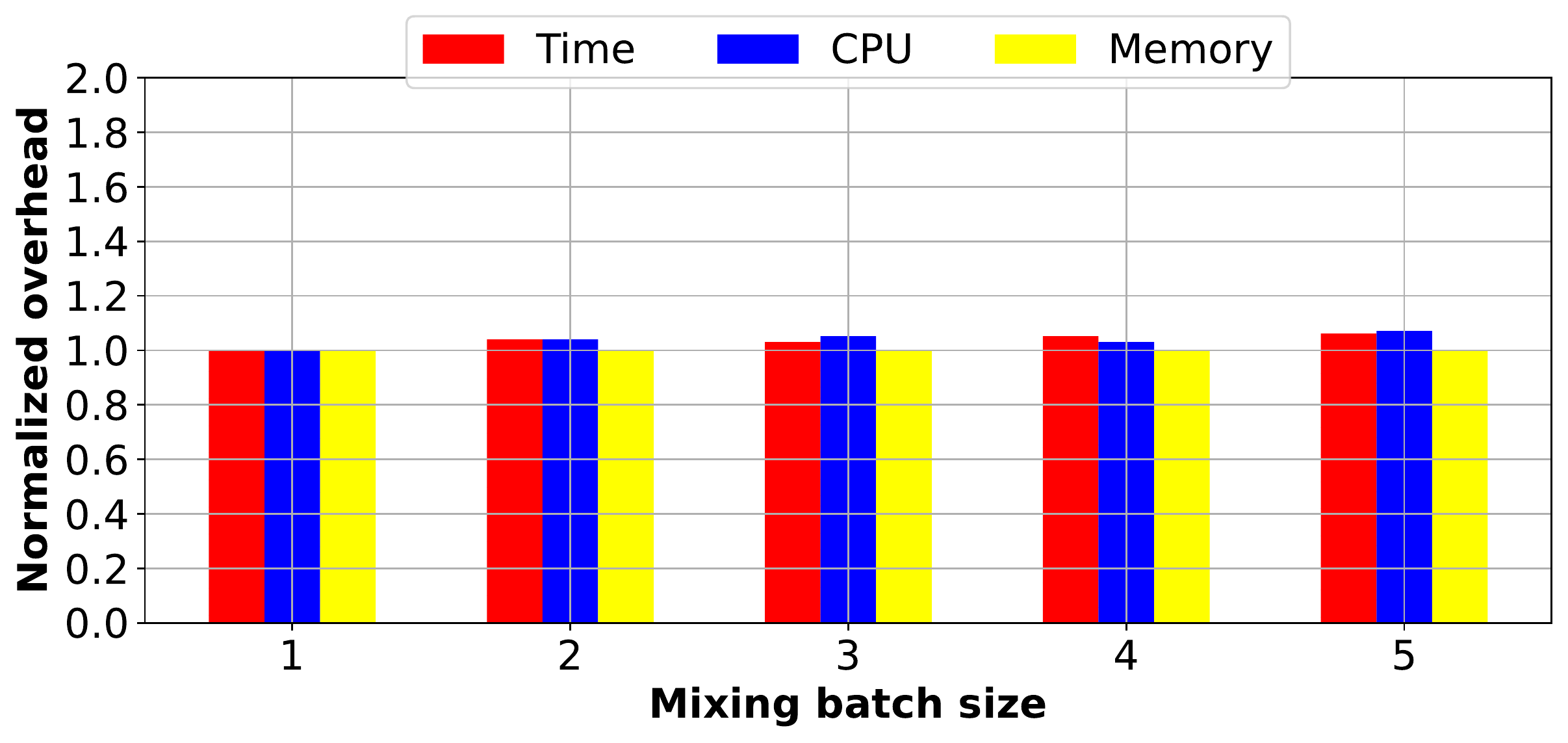}
 \vspace{-0.3cm}
 \caption{Normalized overheads of the parallel \mix mode for different batch sizes} 
 \label{Figure:mixing_parallel_overhead}
 \vspace{-0.7cm}
\end{figure}

\noindent\textbf{Effect of anonymization on overhead:} In Figure \ref{Figure:mixing_overhead} and Figure \ref{Figure:mixing_parallel_overhead}, we present the overhead introduced by mixing computations for different batch sizes. A batch size equal to one indicates that computations are executed without anonymization. As a result, there is no additional overhead in this case. For the sequential \mix mode (Figure \ref{Figure:mixing_overhead}) both the time and CPU overheads show the same trend and increase as the mixing batch size increases. The time overhead is due to the execution of multiple computation graphs in a serialized manner (as shown in Figure \ref{Figure:mix}), whereas the CPU overhead is due to context switching among the nodes of different graphs. On the other hand, Figure \ref{Figure:mixing_parallel_overhead} shows the time and CPU overheads of the parallel \mix mode, which are slightly higher than executing individual computations without anonymization. The time and CPU overheads are due to the fact that each thread needs to select a node of the computation graph, which has not been executed yet by other threads. 

Unlike the \solOne mode, the \mix mode does not add ``fake nodes'' to computation graphs or padding to inputs and outputs to anonymize computations. Because of that, there is no extra memory overhead introduced, thus the memory usage remains the same as the mixing batch size increases. 

\noindent\textbf{Required time to anonymize a computation graph:} In Table \ref{tab:mix_conversation_time}, we present the time required to anonymize a computation graph using the \mix mode for different batch sizes. 
Unlike the \solOne mode, the anonymization time is in the range of a few tens of microseconds, since the \mix mode only needs to select which node of a computation graph to execute instead of trying to conceal the features of computation graphs. 

\begin{table}%
\centering
\caption{Required time to anonymize a computation graph using the \mix mode}
\vspace{-0.2cm}
\resizebox{0.65\columnwidth}{!}{%
\label{tab:mix_conversation_time}
\centering
\begin{tabular}{|c|c|}
\hline
\multicolumn{1}{|l|}{\textbf{\begin{tabular}[c]{@{}c@{}}Mixing batch size\end{tabular}}} & \textbf{\begin{tabular}[c]{@{}c@{}}Required time to anonymize (ms) \end{tabular}}  \\ \hline
2 &  0.009  \\ \hline
3 &  0.011 \\ \hline
4 &  0.019 \\ \hline
5 &  0.020  \\ \hline

\end{tabular}
}
\vspace{-0.6cm}
\end{table}

\subsubsection{\hybrid mode}

\begin{figure*}[!th]
	\centering
	\begin{subfigure}{0.3\textwidth}
		\centering
		\includegraphics[scale=0.2]{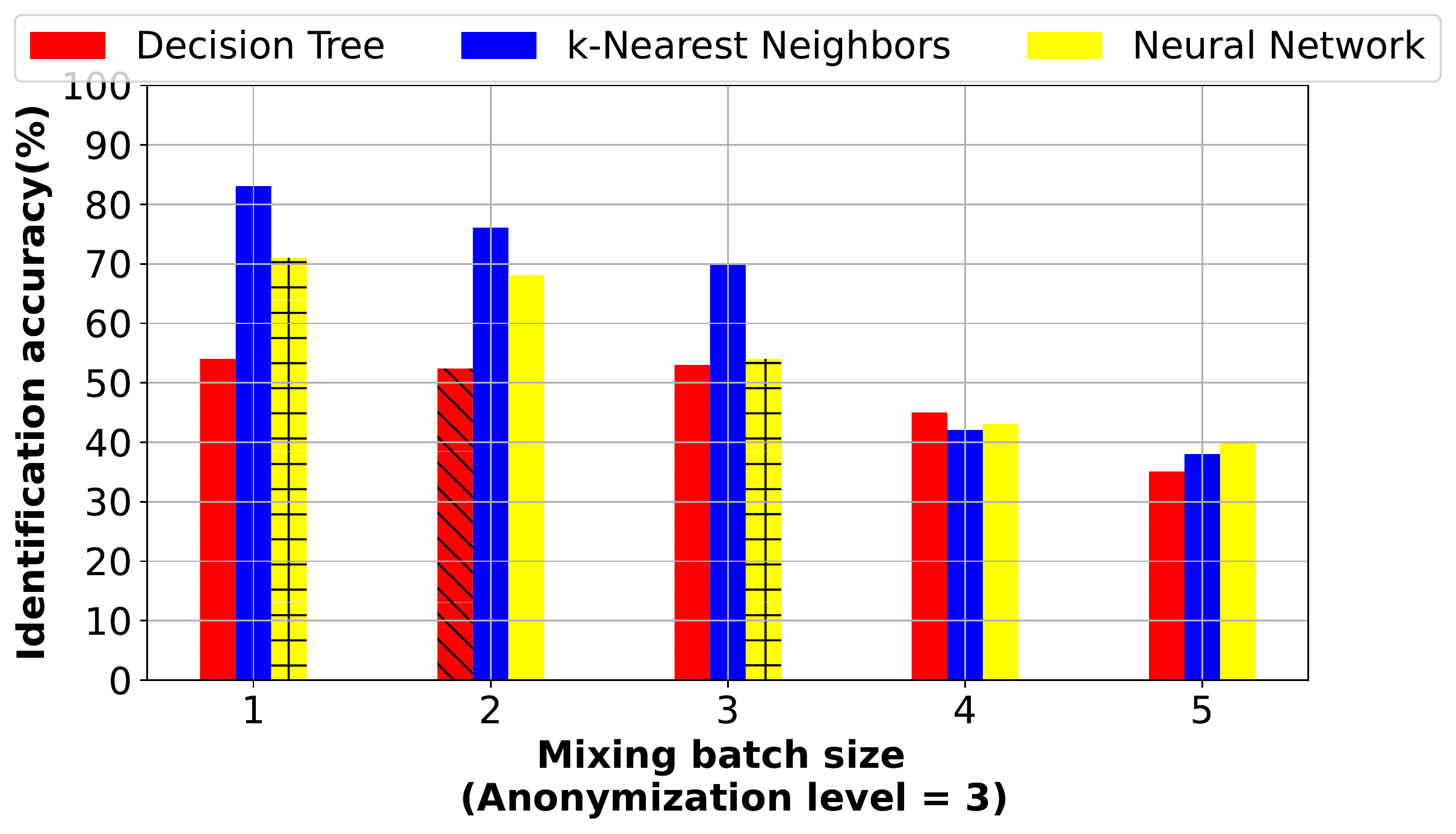}
        \vspace{-0.15cm}
		\caption{Synthetic dataset.}
		\label{Figure:hybrid_syn_acc}
	\end{subfigure} \hfil
	\begin{subfigure}{0.3\textwidth}
	    \centering
	    \includegraphics[scale=0.2]{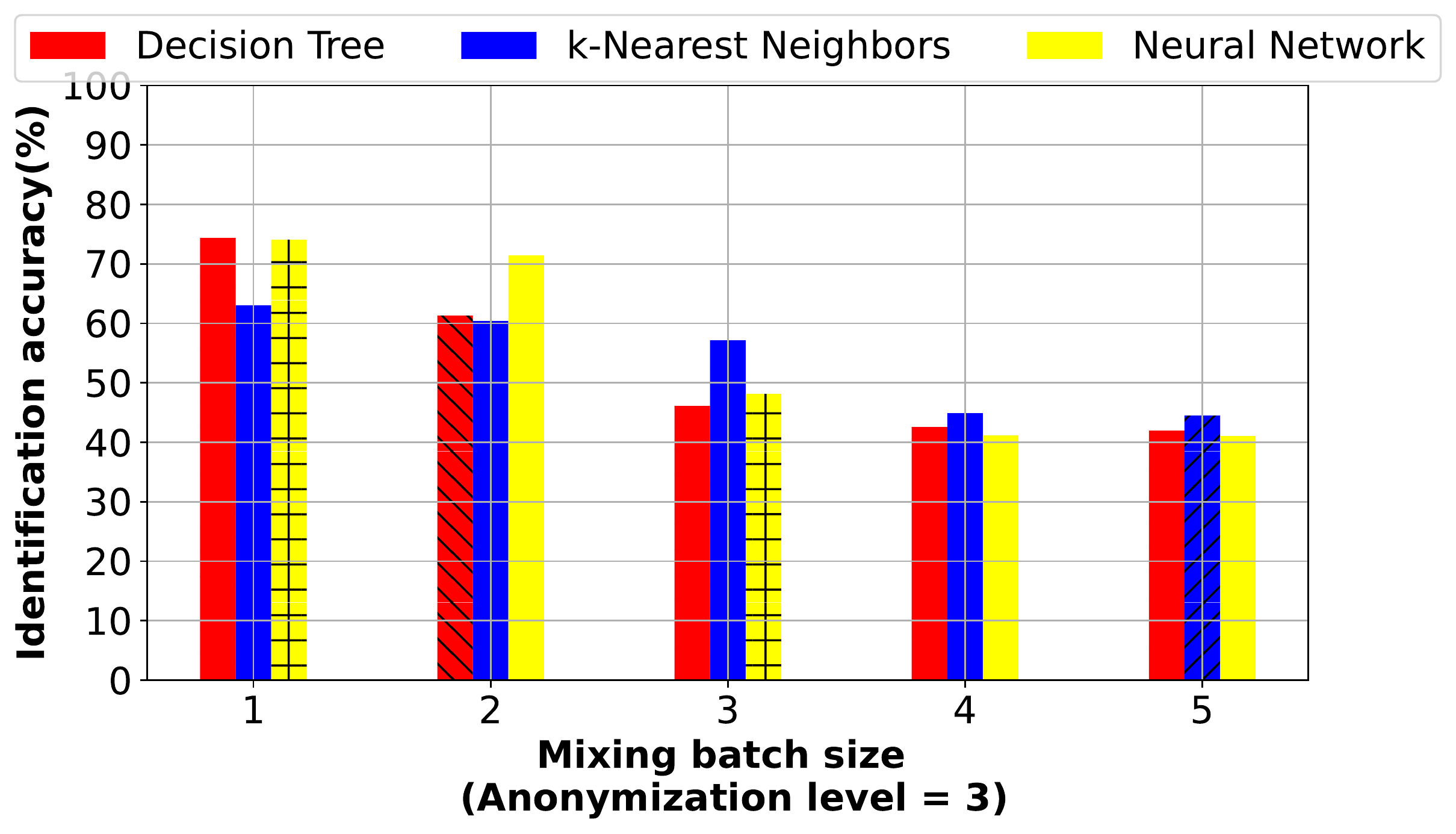}
         \vspace{-0.15cm}
	    \caption{ML/AI dataset}
	    \label{Figure:hybrid_ml_acc}
	\end{subfigure}\hfil
	\begin{subfigure}{0.3\textwidth}
	    \centering
	    \includegraphics[scale=0.2]{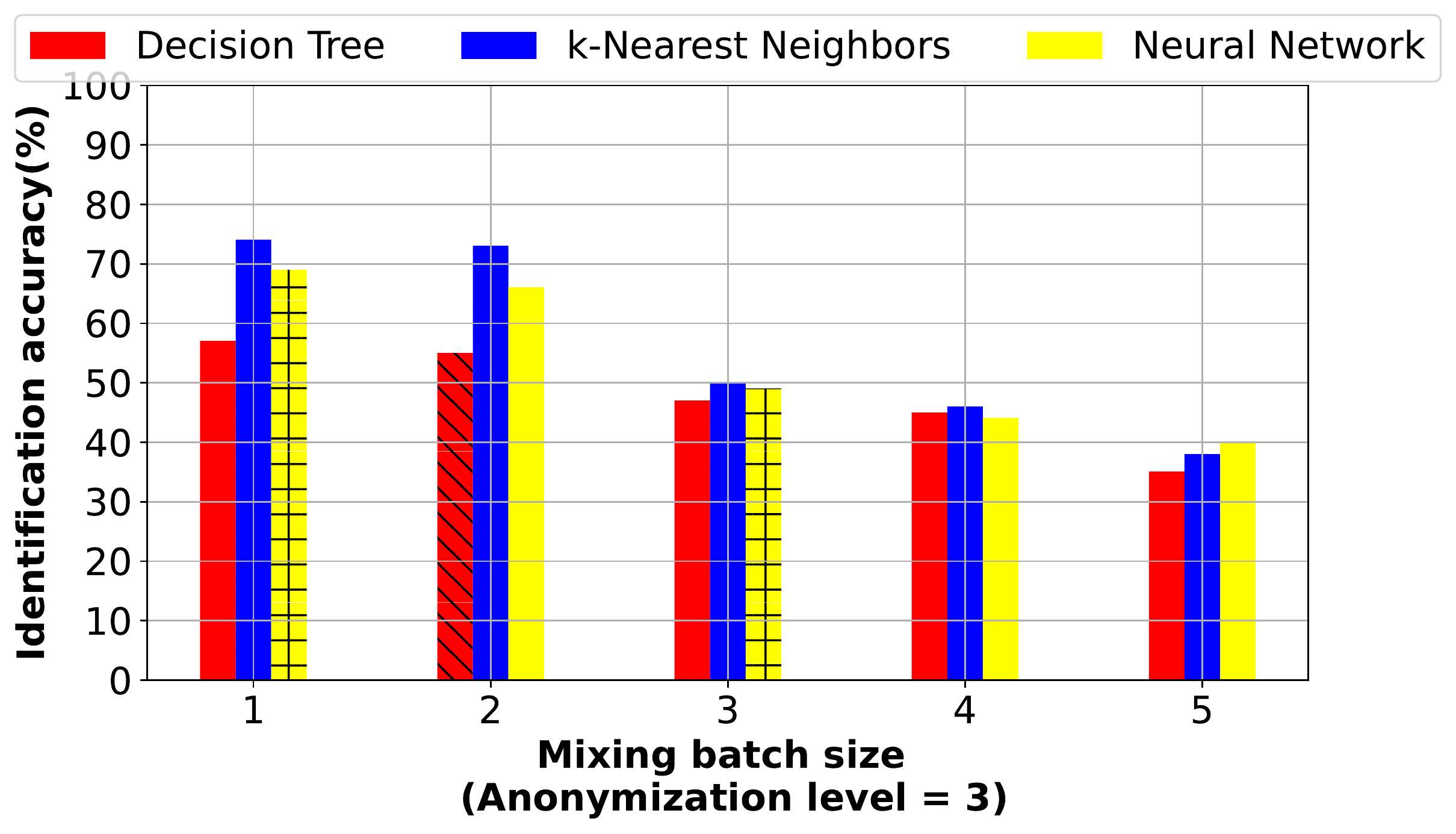}
         \vspace{-0.15cm}
	    \caption{Alibaba cloud dataset.}
	    \label{Figure:hybrid_alibaba_acc}
	\end{subfigure}\hfil
	\vspace{-0.2cm}
	\caption{Identification accuracy of different attack models for different batch sizes using the \hybrid mode.}
	\label{figure:hybrid_acc_batch}
	\vspace{-0.4cm}
\end{figure*}

\noindent\textbf{Effect of anonymization on accuracy:} In Figure \ref{figure:hybrid_acc_batch}, we present results on the accuracy of the \hybrid mode. We first anonymize computation graphs using the \solOne mode (anonymization level equal to 3), and then these anonymized graphs are mixed together using the parallel \mix mode. Our results indicate that the \hybrid mode is more effective in terms of impairing the computation identification ability of attack models as compared to the \solOne mode for the same anonymization level and the \mix mode for the same batch sizes. We would like to note that the results of the \hybrid mode demonstrate the same trend for different anonymization levels as well as for both the sequential and the parallel \mix modes. 

\noindent\textbf{Effect of anonymization on overhead and required time to anonymize computation graphs:} Since the \hybrid mode combines both the \solOne and the \mix modes, the required time to anonymize a computation graph is the summation of the required time to anonymize the graph through the \solOne and \mix modes. In the same manner, the overhead to execute a computation graph due to anonymization is the combined overhead of both modes.

\subsubsection{Adaptive Attack Models} 
\begin{figure*}[!th]
\vspace{-0.1cm}
	\centering
	\begin{subfigure}{0.3\textwidth}
		\centering
		\includegraphics[scale=0.2]{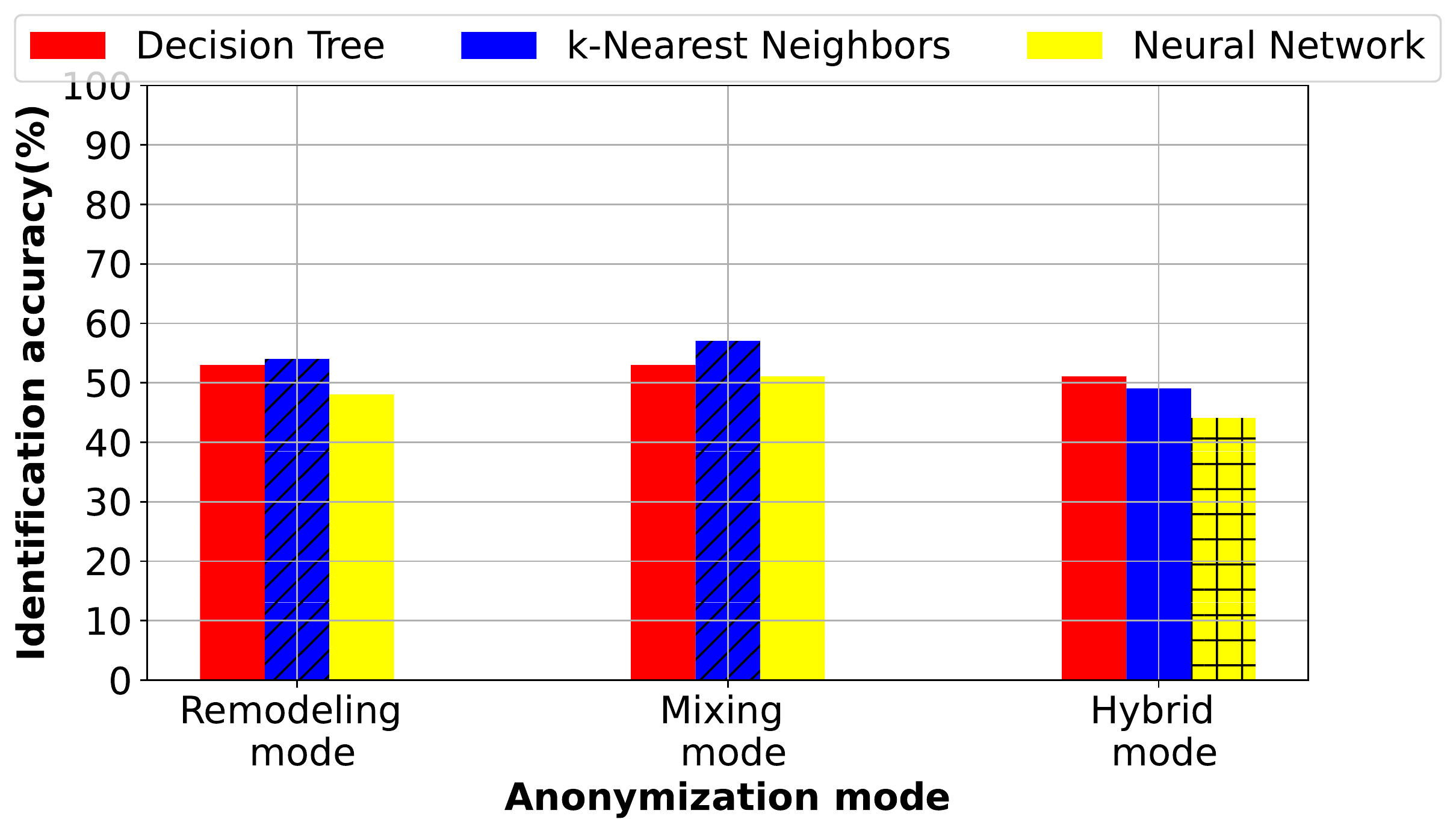}
         \vspace{-0.15cm}
		\caption{Synthetic dataset.}
		\label{Figure:adaptive_syn_acc}
	\end{subfigure} \hfil
	\begin{subfigure}{0.3\textwidth}
	    \centering
	    \includegraphics[scale=0.2]{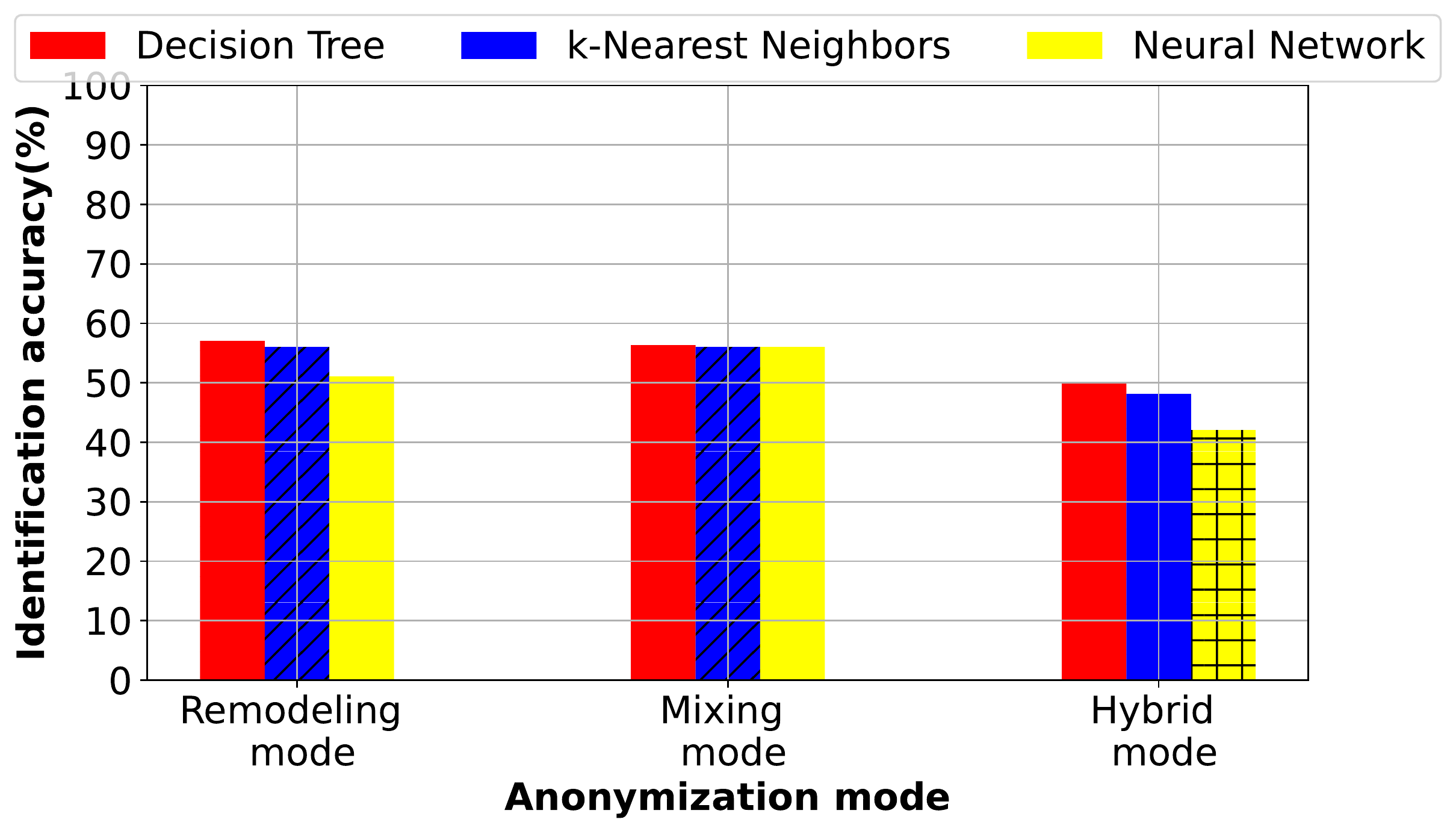}
         \vspace{-0.15cm}
	    \caption{ML/AI dataset}
	    \label{Figure:adaptive_ml_acc}
	\end{subfigure}\hfil
	\begin{subfigure}{0.3\textwidth}
	    \centering
	    \includegraphics[scale=0.2]{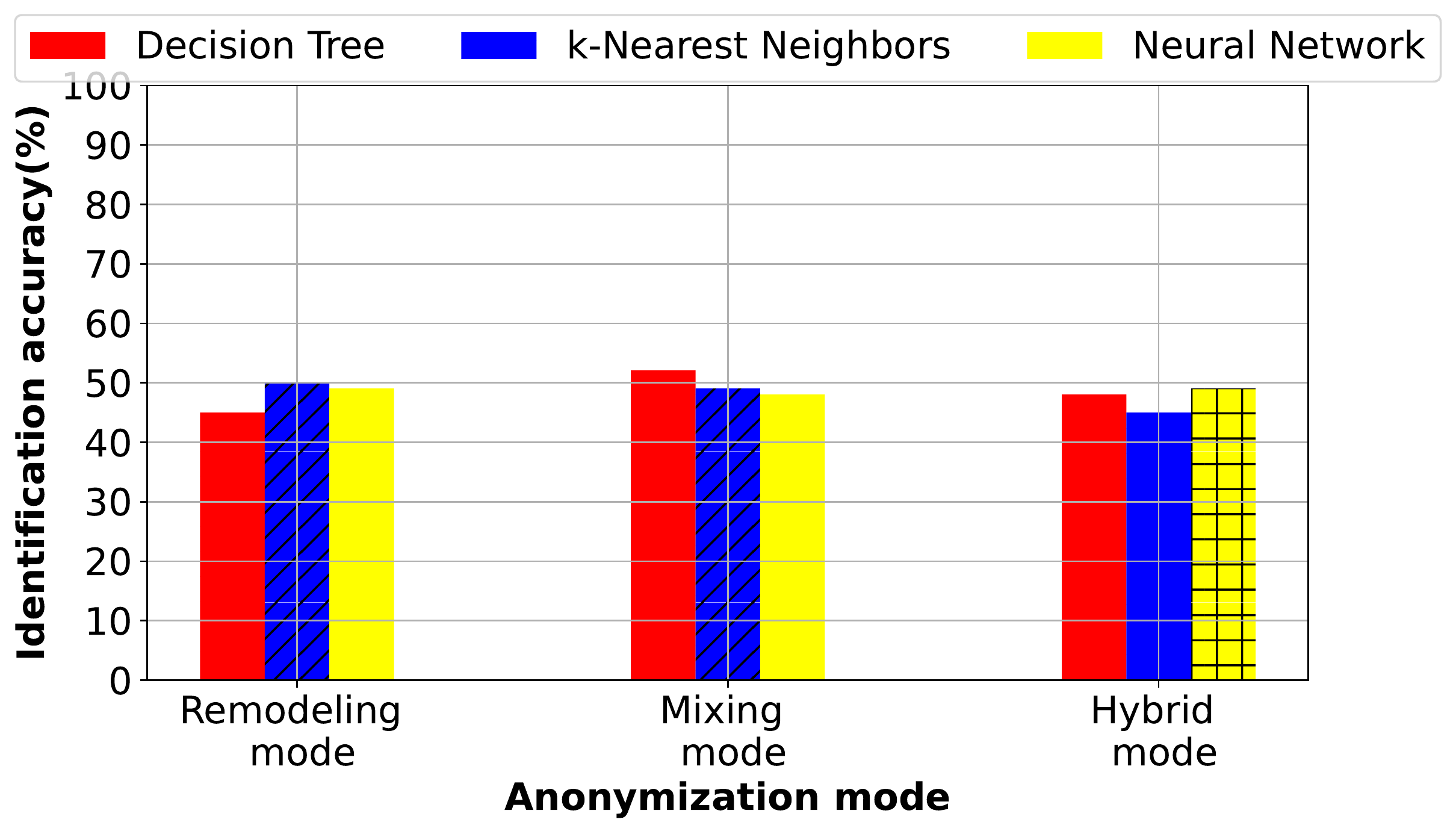}
         \vspace{-0.15cm}
	    \caption{Alibaba cloud dataset.}
	    \label{Figure:adaptive_alibaba_acc}
	\end{subfigure}\hfil
	\vspace{-0.25cm}
	\caption{Performance of different adaptive attack models.}
	\label{figure:adaptive_acc}
	\vspace{-0.75cm}
\end{figure*}

In Figure~\ref{figure:adaptive_acc}, we present the accuracy of adaptive attack models, which are trained based on both anonymized and without anonymization computation graph data. Our results indicate that the accuracy of adaptive attack models remains within the range of 44\%-57\% for all anonynization modes and datasets. In other words, \sol can effectively anonymize computations even when attackers realize that anonymization techniques are used. This is due to the fact that the different modes of \sol make computation execution unpredictable by introducing randomness to the computation process, making computations ``look alike'', and combining the execution of independent computations. 


\section {Discussion}
\label{sec:discussion}




\noindent\textbf{Overhead comparison between the \solOne and \mix modes:} The results of Section \ref{sec:eval} demonstrate that the \mix mode incurs lower time and CPU overheads than the \solOne mode in order to reduce the accuracy of attack models by the same extent, while also incurring no memory overhead. In general, our results show that the accuracy to overhead ratio is lower for the \mix mode than the \solOne mode. The extra overhead of the \solOne mode is due to concealing the features of computations by introducing additional components to the computation process, such as ``fake'' nodes and padded inputs and outputs. On the other hand, the \mix mode does not introduce additional computational components, since nodes from multiple computation graphs are mixed together. 

\noindent\textbf{Mode selection:} When there are no concurrent computations that need to be executed or the same computation needs to be executed multiple times, the \solOne mode should be used. 
When there are multiple independent computations, which can be executed simultaneously, then the \mix mode can provide anonymization with relatively less overhead than the \solOne mode. The \hybrid mode can also be used in such scenarios, which will provide stronger anonymization at the cost of higher overhead. As the results of Table \ref{tab:solOne_conversation_time} indicate, the \solOne mode may not be able to meet the constraints of delay-sensitive computations, if they are anonymized on the fly. In such cases, we can use the \solOne mode with the option of pre-anonymized computations or the \mix mode.



\noindent\textbf{How to apply \sol to white box and black box computations:} Computations can be divided into two categories based on whether we have access to the internal structure (graph) or code of the computation: white box and black box computations. In the case of white box computations, since the graphs of the computations are known to \sol, both the \solOne mode or the \mix mode can be applied for their anonymization. Without being aware of the computation graph structure (black box computations), although it will not be possible to use the \mix mode, we can still apply the \solOne mode. To achieve that, for example, we can add ``fake'' nodes at the beginning and the end of computations to manipulate memory and CPU usage as well as use these ``fake'' nodes to add or remove input and output padding.

\section{Related Work}
\label{sec:back}



\subsection {Adversarial Machine Learning} 

Adversarial machine learning refers to techniques that allow an attacker to obtain information about machine learning models and take advantage of this information to confuse models and lead them to making wrong predictions \cite{barreno2006can}. Papernot \textit{et al} demonstrated that adversarial inputs generated for a specific model can be used for a different target model as long as both models are trained to carry out the same task, even though their architectures and training sets may be different~\cite{papernot2016transferability}. This transferability of adversarial input data can lead to black box attacks, where attackers do not need to possess detailed information about target models \cite{papernot2017practical}. Brendel \textit{et al} proceeded one step further and claimed that an attacker can perform black box attacks against a victim model only based on the final model output \cite{brendel2017decision}. 

Szegedy \textit{et al} demonstrated that the prediction error of a trained deep learning image classification model can be maximized by adding imperceptible perturbations to input images \cite{szegedy2013intriguing}. Deepfool \cite{moosavi2016deepfool} was proposed by Moosavi-Dezfooli \textit{et al} to generate adversarial input data by introducing minimal perturbations to confuse trained models. In some cases, it is possible to change the prediction outcome of a well-trained image classification model by carefully modifying a single pixel of an input image \cite{su2019one}. Such an adversarial strategy is not exclusive to high dimensional data, such as image and speech, but to other domains as well. Li \textit{et al} presented ConAML \cite{li2021conaml} to generate adversarial samples for cyber-physical systems. Finally, an attack based on adversarial machine learning was proposed in the context of Internet of Things (IoT), where a fraction of the available IoT devices needs to be manipulated during the data fusion/aggregation process~\cite{luo2020adversarial}.

\subsection {Privacy-Preserving Computation}


$VC^{3}$ \cite{schuster2015vc3} was proposed to execute MapReduce \cite{dean2008mapreduce} jobs on an untrusted cloud while preserving the confidentiality of jobs (computations) by keeping the code and data secret. Dinh \textit{et al} proposed $M^{2}R$ \cite{dinh2015m2r} to improve the privacy of sensitive MapReduce computation tasks by hiding data access patterns of the whole distributed computing framework instead of securing each computing node individually, since the exchanges of the input and output data among the computing units can reveal the type or identity of ongoing computations. A multi-party secure computation framework was proposed by Mazloom \textit{et al}, where certain information can be learned about computations while preserving the differential privacy of the input of computation graphs, such as histograms, PageRank, and matrix factorization~\cite{mazloom2018secure}. Ohrimenko \textit{et al} showed that observing encrypted intermediate data among MapReduce units may leak sensitive information about computations \cite{ohrimenko2015observing}. 

One way to preserve the confidentiality of computations in an untrusted/compromised environment is the use of homomorphic encryption \cite{gentry2009fully}. Frameworks based on homomorphic encryption, such as CryptDB \cite{popa2011cryptdb}, MrCrypt \cite{tetali2013mrcrypt}, Monomi \cite{10.14778/2535573.2488336}, and AUTOCRYPT \cite{tople2013autocrypt}, \cite{nikolaenko2013privacy, nikolaenko2013privacy2}, have been proposed to enable the privacy of computations. However, homomorphic encryption is expensive and remains in general impractical for computations on large data.

Trusted hardware has also been used for the execution of tasks in a privacy-preserving manner. Cipherbase \cite{arasu2013orthogonal}, TrustedDB \cite{bajaj2013trusteddb}, Monomi \cite{10.14778/2535573.2488336} are proposed frameworks that use trusted hardware. Furthermore, GraphSC \cite{nayak2015graphsc} and ObliVM \cite{liu2015oblivm} are programming paradigms that provide secure computations and can be integrated into computation frameworks, such as MapReduce, GraphLab \cite{low2012distributed}, and Spark \cite{zaharia2010spark}. Finally, research has been conducted on anonymizing social graphs \cite{feder2008anonymizing, liu2008towards, sala2011sharing, zhang2017towards}. Nevertheless, social graphs may have different characteristics than computation graphs.


\subsection {Side Channel Attacks and Protection}

With the advancement of multi-tenant clouds, the security and privacy of users' sensitive data and computations have become important concerns. There have been several channel attacks studied for multi-tenant cloud platforms. Zhang \textit{et al} demonstrated that an attacker can infer various information about a victim's web applications on commercial Platform-as-a-Service clouds using a cache-based side channel attack \cite{zhang2014cross}. Pessl \textit{et al} presented a side channel attack, called DRAMA \cite{pessl2016drama}, based on the mapping of the memory address to DRAM. Varadarajan \textit{et al} demonstrated that even if there have been improvements to prevent side channel attacks in public multi-tenant clouds and some earlier attacks have become ineffective, it is still possible that an attacker can be co-located with a targeted victim, thus observing and collecting information leaked by the victim's VM instance \cite{varadarajan2015placement}.

CacheAudit \cite{doychev2015cacheaudit} was proposed to analyze binary executable programs in order to automatically assess the amount of information leaked by cache access and timing patterns. STEALTHMEM \cite{kim2012stealthmem}, CloudRadar \cite{zhang2016cloudradar}, and CATalyst \cite{liu2016catalyst} are frameworks that have been proposed to prevent cached-based side channel attacks in multi-tenant clouds. Finally, D{\'e}j{\'a} Vu \cite{chen2017detecting} is a framework that detects and prevents an attacker from observing side channel information in a fine-grained manner. 
\section {Conclusion and Future Work}
\label{sec:conclusion}

In this paper, we explored the design space of anonymizing offloaded computations through the realization of \sol, a framework that employs practical mechanisms 
to conceal features/properties of executed computations. 
Our evaluation results demonstrated that \sol can effectively anonymize computations and hamper the ability of attackers to identify executed computations even when attackers realize that anonymization techniques are used. In the future, we will explore the following directions: (i) theoretically formulate the operation of \sol; (ii) expand the evaluation of \sol to include additional scenarios and applications; and (iii) focus on the case of black box computations, where the structure of computation graphs may be unknown.

\section*{Acknowledgements}

This work is partially supported by the National Science Foundation (awards CNS-2104700, CNS-2306685, CNS-2016714, and CBET-2124918) and ACM SIGMOBILE.

\bibliographystyle{IEEEtran}
\bibliography{refs}

\end{document}